# Deep Reinforcement Learning Approach to QoS Aware Load Balancing in 5G Cellular Networks under User Mobility and Observation Uncertainty

**M. Eskandarpour, H. Soleimani**

School of Electrical Engineering, Iran University of Science & Technology (IUST), Tehran, Iran

**(Corresponding Author: hsoleimani@iust.ac.ir)**

*Abstract*— **Efficient mobility management and load balancing are essential for maintaining Quality of Service (QoS) in dense and highly dynamic 5G radio access networks. This paper proposes a Proximal Policy Optimization (PPO)-based deep reinforcement learning framework for autonomous QoS-aware load balancing through adaptive Cell Individual Offset (CIO) control. The problem is formulated as a Markov Decision Process in which the agent periodically adjusts bounded CIO values to influence user association and handover behavior. To capture the multi-objective nature of 5G load balancing, the reward function jointly considers aggregate throughput, latency, jitter, packet-loss ratio, Jain's fairness index, and handover rate. The proposed PPO agent employs an actor–critic neural architecture and is trained in a lightweight Python-based simulation environment that includes Gauss–Markov user mobility, stochastic traffic dynamics, channel variation, and noisy observations. Simulation results over more than 500 training episodes demonstrate that the proposed method achieves stable convergence and improves network performance by increasing throughput and fairness while reducing latency, jitter, packet loss, and unnecessary handovers. Comparative evaluations against A3, ReBuHa, and clipped double Q-learning (CDQL) baselines show that PPO provides the best overall QoS trade-off under the evaluated mobility, user-density, and noisy-observation settings. These results indicate that advantage-based policy optimization with clipped updates is a promising approach for robust and adaptive load balancing in next-generation 5G radio access networks.**

## I. INTRODUCTION

The rapid growth of mobile services [1] and the increasing demand for bandwidth-intensive [2], [3] and latency-sensitive applications [4] have significantly transformed modern wireless communication systems [5]. Emerging services such as ultra-high-definition video streaming, augmented and virtual reality (AR/VR), autonomous driving [6], and massive Internet of Things (IoT) deployments impose stringent requirements on cellular infrastructure in terms of capacity, latency, reliability, and scalability [7]. To satisfy these requirements, fifth-generation (5G) networks must support high data rates, low latency, reliable connectivity [8], and intelligent resource allocation [9] across dense and highly dynamic Radio Access Networks (RANs) [10]. In this context, load balancing plays a critical role in maintaining Quality of Service (QoS), improving spectral efficiency, preventing cell congestion, and preserving user experience under heterogeneous traffic demands and mobility patterns [11], [12].

Conventional mobility management mechanisms, such as A3-based handover, rely mainly on downlink radio measurements, including Reference Signal Received Power (RSRP), and compare them against fixed thresholds [13], [14]. These mechanisms typically use predefined parameters such as hysteresis margin and Time-To-Trigger (TTT) to avoid unnecessary handovers [15]. Although such rule-based methods are simple and widely deployable, their static nature limits their ability to adapt to rapid variations in user mobility, traffic distribution, and cell congestion. Moreover, conventional handover mechanisms mainly focus on radio-signal strength and often do not explicitly account for end-to-end QoS indicators such as latency, jitter, packet loss, or system-level fairness [16], [17]. Resource-aware heuristic methods, such as ReBuHa, improve upon purely signal-based handover by incorporating load-related indicators such as Resource Block Utilization (RBU) [18]. However, these approaches remain rule-based and may become ineffective under high mobility, stochastic traffic, and noisy or partially observable network conditions [19].

These limitations have motivated the use of Reinforcement Learning (RL) for self-optimizing wireless networks [20]. In RL-based network control, an agent learns sequential decision-making policies through interaction with the environment and feedback from performance-driven rewards [21]. Among deep RL algorithms, Proximal Policy Optimization (PPO) has attracted considerable attention due to its stable policy updates, compatibility with continuous and discrete action spaces, and strong empirical performance in complex control problems [22], [23]. PPO uses a clipped surrogate objective to restrict large policy updates, advantage-based learning to improve sample efficiency, and entropy regularization to encourage exploration. These characteristics make PPO suitable for non-stationary wireless environments in which the controller must balance multiple and sometimes conflicting objectives under uncertainty.

In this paper, we propose a PPO-based deep reinforcement learning framework for QoS-aware load balancing in dense 5G RANs. The control problem is formulated as a Markov Decision Process (MDP), where the agent periodically adjusts Cell Individual Offset (CIO) values to influence user association and handover behavior. The state includes aggregated radio, load, and QoS indicators, such as cell utilization, RSRP/CQI-related measurements, latency, jitter, packet-loss ratio, throughput, and recent handover activity. The action is defined as a bounded CIO adjustment vector, allowing the controller to gradually shift cell attractiveness while avoiding abrupt changes that may lead to ping-pong handovers. The reward function combines multiple QoS objectives, including aggregate throughput, latency, jitter, packet-loss ratio, Jain's fairness index, and handover count. This multi-objective formulation enables the learned policy to balance network efficiency, fairness, and mobility stability.

To improve realism, the proposed framework is evaluated in a Python-based simulation environment that includes stochastic user mobility, wireless channel variation, traffic dynamics, and noisy observations. User mobility is modeled using the Gauss–Markov process to capture temporally correlated movement patterns, while measurement uncertainty is introduced through controlled noise on key reported indicators. These assumptions allow the agent to be trained and tested under conditions that more closely resemble practical RAN operation, where network telemetry is imperfect, delayed, and partially observable.

The proposed PPO controller is compared with two classical baselines, A3 and ReBuHa, as well as a learning-based clipped double Q-learning (CDQL) baseline. All methods are evaluated under the same mobility, traffic, and observation-noise settings to ensure a fair comparison. Simulation results over more than 500 training episodes and additional stress tests with increasing user density show that the proposed PPO-based controller achieves higher throughput and fairness while reducing latency, jitter, packet loss, and unnecessary handovers. Compared with rule-based schemes and the CDQL baseline, PPO provides smoother convergence and a stronger overall QoS trade-off under dynamic load and mobility conditions.

The main contributions of this paper are summarized as follows:

1. **PPO-based CIO control for 5G load balancing:** We develop an actor–critic PPO controller that adaptively adjusts CIO values to influence user association and handover behavior under mobility.
2. **Multi-objective QoS-aware reward design:** We formulate a composite reward that jointly considers throughput, latency, jitter, packet-loss ratio, fairness, and handover activity, enabling the agent to balance efficiency and stability.
3. **Mobility- and noisy-observation-aware evaluation:** We incorporate Gauss–Markov user mobility and stochastic observation noise to evaluate the behavior of the learned policy under practical measurement uncertainty.
4. **End-to-end Python-based simulation framework:** The network environment, mobility model, observation model, PPO training pipeline, and evaluation procedure are implemented in Python, supporting reproducibility and rapid experimentation.
5. **Comprehensive baseline comparison:** We benchmark the proposed PPO method against A3, ReBuHa, and CDQL, demonstrating improved QoS performance and more stable convergence across the considered scenarios.

The remainder of this paper is organized as follows. Section II reviews related work on RAN load balancing, handover optimization, and reinforcement learning-based network control. Section III presents the system model and MDP formulation. Section IV describes the proposed PPO-based methodology, including the actor–critic architecture, advantage estimation, reward processing, and training procedure. Section V presents the simulation setup and performance evaluation. Section VI discusses the results, limitations, and practical considerations. Finally, Section VII concludes the paper and outlines future research directions.

## II. Related Work

Mobility management and load balancing are long-standing challenges in cellular radio access networks, but they become more difficult in dense 5G deployments because user association decisions are affected by several interacting factors. A user may receive the strongest signal from one base station, while another nearby cell may provide better service because it has lower load, shorter queues, or more available radio resources. Therefore, an effective load-balancing mechanism should not only consider radio measurements, but also account for resource utilization, QoS degradation, and mobility stability.

Classical handover mechanisms in LTE and 5G networks are usually based on radio measurements such as Reference Signal Received Power (RSRP), Reference Signal Received Quality (RSRQ), or SINR. In event-based handover procedures, such as the A3 event, a handover is triggered when the target cell becomes better than the serving cell by a predefined offset for a sufficient Time-To-Trigger duration. These methods are simple, standardized, and widely deployable, but their performance strongly depends on the tuning of parameters such as hysteresis, Time-To-Trigger, and Cell Individual Offset (CIO). Survey studies on handover management show that static handover parameters may lead to handover failures, ping-pong events, or inefficient load distribution when user mobility, traffic demand, and interference conditions change rapidly [24], [25]. Recent studies also emphasize that traditional handover and load-balancing mechanisms are increasingly challenged by ultra-dense 5G/6G environments, where small cells, heterogeneous services, and dynamic user behavior make fixed-threshold control less effective [67].

Load balancing is often handled through handover-parameter optimization. In this approach, the network does not directly force users to move between cells; instead, it adjusts handover-related parameters so that users near cell boundaries are more likely to associate with less congested cells. CIO is particularly important in this context because it biases the effective attractiveness of a cell without changing the physical radio channel. A positive CIO can attract more users to a cell, while a negative CIO can discourage association with an overloaded cell. However, overly aggressive CIO adjustment may cause unstable association behavior, especially near cell edges. Therefore, CIO-based load balancing must balance congestion relief against mobility stability.

Several machine learning and reinforcement learning approaches have been proposed to improve handover and mobility load balancing. A Q-learning-based handover approach for 5G networks was proposed in [26], where the learning agent improves mobility decisions by adapting to the observed network state. Deep learning has also been used for predictive handover. For example, in [27], SINR variation is used to predict radio-link failure probability and support target-cell selection. For ultra-dense networks, an RL-based mobility load-balancing framework was introduced in [28]. That method uses a two-layer structure in which cells are first clustered and then handover parameters are optimized inside each cluster based on PRB utilization. Another RL-based load-balancing method was proposed in [29], where the objective is to improve instantaneous network throughput. These works show that learning-based mobility control can outperform static rule-based handover schemes. However, many of them focus mainly on signal quality, PRB utilization, or throughput, while broader user-experienced QoS metrics such as latency, jitter, packet-loss ratio, fairness, and handover stability are not always jointly optimized.

More recent studies have also considered learning-based mobility load balancing in architectures closer to O-RAN. For example, a 2023 study proposed an O-RAN-compliant Q-learning algorithm that dynamically adapts the handover offset between neighboring cells to address cell overload while preserving the required QoS of connected users [68]. This is closely related to CIO-based mobility load balancing, because it uses learning to adjust handover bias rather than relying only on fixed thresholds. However, Q-learning methods usually require discrete action spaces, which can limit the resolution of CIO adjustment when the controller must manage several cells jointly. In contrast, the present work uses PPO to generate continuous bounded CIO adjustments, allowing smoother control and avoiding coarse action discretization.

Reinforcement learning has also been investigated more broadly for communication load balancing. A recent review highlights that load balancing can be addressed by rule-based, optimization-based, and learning-based methods, and it identifies CIO, hysteresis, and Time-To-Trigger as central parameters in handover-based load balancing [69]. This supports the motivation for using CIO as the control variable in the proposed framework. At the same time, the review also points out that communication load balancing remains difficult because the controller must react to time-varying demand, user mobility, and imperfect observations. These challenges are directly relevant to the present work, where PPO is trained under mobility, channel variation, traffic dynamics, and noisy measurements.

Another related research direction is traffic steering and load balancing in O-RAN. O-RAN introduces the near-real-time RAN Intelligent Controller, where xApps can perform closed-loop control using network telemetry and policy feedback. A 2024 study on policy-based traffic steering and load balancing in O-RAN-based V2N communication shows that O-RAN can support dynamic policy control for vehicular services and load management through RIC-based applications [70]. This line of work is important because it connects load-balancing algorithms with practical deployment architectures. However, many O-RAN traffic-steering studies rely on rule-based or policy-based control rather than end-to-end reinforcement learning with continuous CIO optimization. The proposed PPO-based controller is therefore aligned with the O-RAN vision, but focuses more specifically on learning a smooth CIO-control policy from multi-KPI QoS feedback.

Beyond handover and CIO optimization, reinforcement learning has been widely applied to radio resource management. UAV-assisted vehicular downlink resource allocation was studied in [30], while PPO-based joint power and bandwidth allocation for V2V communication was investigated in [31]. A centralized hierarchical DRL framework for relay selection and power allocation in mmWave vehicular networks was proposed in [32], and a GNN-based method for wireless power control was presented in [33]. These studies demonstrate the flexibility of learning-based control in wireless networks, especially when decisions must be made under mobility and interference. However, their main control variables are power, bandwidth, relay selection, or graph-based resource allocation, not CIO-based user association and handover bias control. Therefore, they are useful for positioning the proposed work within the broader DRL-based RAN optimization literature, but they do not directly address the same problem.

Power control and joint resource optimization have also received significant attention in cellular and mmWave networks. Reinforcement learning has been used for joint power and channel allocation [34], joint handover control and power allocation with multi-agent PPO [35], user association and power control in ultra-dense mmWave small cells [37], and DRL-based spectrum–power allocation with spectral-efficiency, energy-efficiency, and fairness objectives [38]. Other studies have considered mmWave small-cell resource management [36], power control with caching for vehicular video delivery [39], uplink resource allocation and beamforming [40]–[44], and joint beamforming, power control, and interference coordination [45]–[49]. These works are important because they show how learning-based optimization can handle multiple coupled radio-control variables. Nevertheless, power-control problems are different from CIO-based mobility load balancing. Power control directly changes the physical-layer transmission behavior, whereas CIO control modifies association and handover decisions while keeping compatibility with conventional mobility-management procedures.

Deep learning has further been used for several wireless control and prediction tasks. Examples include online MIMO link adaptation [52], interference avoidance in heterogeneous networks [53], closed-loop power control [54], dynamic multi-channel access [55], automatic modulation recognition [56], deep power control [57]–[59], coordinated mmWave beam prediction [60], channel-to-beamformer mapping [61], antenna-pattern synthesis [62], automated cellular network tuning [63], mmWave beam management and interference coordination [64], and MISO downlink beamforming [65]. These studies show that data-driven methods can learn complex wireless relationships that are difficult to model analytically. However, most of them focus on physical-layer or link-layer optimization and do not directly solve the problem of QoS-aware CIO adjustment under mobility and observation uncertainty.

The work most closely related to the present study is [66], which proposes QoS-aware load balancing using clipped double Q-learning. Both [66] and this paper use a multi-KPI reward structure to improve network performance beyond a single throughput-based objective. However, there are several important differences. First, [66] adopts a value-based CDQL method, while this paper uses a policy-gradient PPO framework. This distinction is important because CIO adjustment is naturally continuous, and PPO can directly produce bounded continuous actions without requiring the CIO space to be discretized. Second, the proposed method explicitly controls user association through CIO adjustment, which makes it compatible with standard handover logic based on cell biasing. Third, the reward design considers throughput, latency, jitter, packet-loss ratio, Jain's fairness index, and handover activity, allowing the controller to balance network efficiency, user-perceived QoS, and mobility stability. Finally, the proposed simulation framework includes mobility, channel variation, stochastic traffic, and noisy observations, which are essential for evaluating whether the learned policy can remain stable under practical RAN uncertainty.

In summary, existing studies have made important progress in handover optimization, mobility load balancing, power control, beam management, and O-RAN traffic steering. However, there is still a need for a learning-based controller that can perform smooth CIO adjustment, jointly optimize multiple QoS metrics, and operate under mobility and imperfect observations. The proposed PPO-based framework addresses this gap by formulating CIO control as a continuous-action reinforcement learning problem and by learning bounded CIO updates that improve load distribution while reducing unnecessary handovers.

## III. System Model and Problem Formulation

This section presents the system model used for QoS-aware load balancing in a 5G downlink cellular network. We first describe the network architecture, radio channel model, traffic and scheduling model, user mobility model, and observation model under measurement uncertainty. We then formulate the load-balancing problem as a Markov Decision Process (MDP), where the reinforcement learning agent dynamically adjusts Cell Individual Offset (CIO) values to influence handover and user association decisions.

The objective of the system model is to capture the main interactions that determine QoS in a dense and mobile 5G radio access network. In such networks, user performance is affected not only by radio signal strength but also by cell load, resource availability, queueing delay, packet loss, and mobility-induced handovers. A UE connected to the strongest cell may still experience poor service if that cell is highly congested. Conversely, associating a UE with a slightly weaker but less loaded neighboring cell may improve latency, packet delivery, and fairness. Therefore, the proposed model jointly considers radio quality, traffic dynamics, scheduling, mobility, and QoS metrics.

The load-balancing controller operates at discrete control intervals. At each decision epoch, it receives aggregated network measurements from base stations, observes the current load and QoS condition, and selects a CIO adjustment action. The updated CIO values bias the handover logic and influence future user-cell associations. The network then evolves according to UE mobility, wireless propagation, traffic arrivals, scheduling decisions, and handover execution. This closed-loop interaction provides the basis for training a deep reinforcement learning agent.

*A. Downlink Cellular Network Model*

We consider a downlink 5G cellular network consisting of $M$ base stations or sectors and $N$ mobile user equipments. The set of base stations is denoted by

$$\Gamma = \{BS_1, BS_2, \dots, BS_M\}, \quad (1)$$

and the set of UEs is denoted by

$$\Psi = \{UE_1, UE_2, \dots, UE_N\}. \quad (2)$$

Each base station operates over a downlink bandwidth of $B$MHz. The available spectrum is divided into $R$ physical resource blocks (PRBs), which are allocated to active UEs by the MAC scheduler. At each time instant, each UE is associated with one serving base station. The serving base station is responsible for downlink data transmission, queue management, and measurement reporting for that UE.

The simulation area is modeled as a bounded two-dimensional region. The position of UE $u$at time $t$ is represented by

$$\mathrm{p}_u(t) = [x_u(t), y_u(t)]^T, \quad (3)$$

and the position of base station $i$ is denoted by

$$\mathrm{p}_{BS_i} = [x_{BS_i}, y_{BS_i}]^T. \quad (4)$$

The Euclidean distance between UE $u$and base station $i$is therefore

$$d_{u,i}(t) = \| \mathrm{p}_u(t) - \mathrm{p}_{BS_i} \|_2. \quad (5)$$

This distance is used to compute path loss and received signal power. As UEs move through the service area, the distances between UEs and base stations change continuously, leading to time-varying received power, SINR, CQI, and handover conditions.

At every control interval $\Delta T$, the network collects radio and QoS measurements. These measurements may include RSRP, RSRQ, SINR, CQI, PRB utilization, throughput, delay, jitter, packet-loss ratio, and handover count. Instead of using instantaneous per-packet information, the controller operates on aggregated per-cell statistics, which is more realistic for a near-real-time load-balancing controller.

*B. Radio Propagation and SINR Model*

The received signal quality between a UE and a base station depends on path loss, shadowing, fading, interference, and thermal noise. The large-scale path loss between UE $u$and base station $i$ is modeled as

$$PL_{u,i}(t) = PL_0 + 10n\log_{10}\left(\frac{d_{u,i}(t)}{d_0}\right), \quad (6)$$

where $PL_0$ is the reference path loss at distance $d_0$, and $n$ is the path-loss exponent. This model captures the average attenuation caused by distance-dependent propagation loss.

To represent slow variations caused by buildings, obstacles, and environmental clutter, log-normal shadowing is included:

$$X_{u,i}^{\mathrm{sh}}(t) \sim \mathcal{N}(0, \sigma_{\mathrm{sh}}^2), \quad (7)$$

where $\sigma_{\mathrm{sh}}$ is the shadowing standard deviation in dB. Small-scale fading is also included to model rapid channel fluctuations caused by multipath propagation. Let $F_{u,i}(t)$ denote the small-scale fading term in dB. The received power from base station $i$at UE $u$can be written as

$$P_{u,i}^{\mathrm{rx}}(t) = P_i^{\mathrm{tx}} + G_i^{\mathrm{ant}} - PL_{u,i}(t) + X_{u,i}^{\mathrm{sh}}(t) + F_{u,i}(t), \quad (8)$$

where $P_i^{\mathrm{tx}}$ is the transmit power of base station $i$, and $G_i^{\mathrm{ant}}$ is its antenna gain. In linear scale, the corresponding channel gain is denoted by $G_{u,i}(t)$. If UE $u$is served by base station $i$, its downlink SINR is given by

$$\mathrm{SINR}_{u,i}(t) = \frac{P_i^{\mathrm{tx}} G_{u,i}(t)}{\sum_{j \neq i} P_j^{\mathrm{tx}} G_{u,j}(t) + N_0 B}, \quad (9)$$

where the denominator includes inter-cell interference from neighboring base stations and thermal noise over bandwidth $B$.

The SINR is then mapped to CQI and modulation-and-coding scheme values. These values determine the achievable spectral efficiency and therefore the amount of data that can be served when PRBs are allocated to the UE.

The achievable downlink rate for UE $u$served by base station $i$can be approximated as

$$R_{u,i}(t) = b_{u,i}(t) B_{\mathrm{RB}} \log_2\left(1 + \mathrm{SINR}_{u,i}(t)\right), \quad (10)$$

where $b_{u,i}(t)$ is the number of PRBs assigned to UE $u$, and $B_{\mathrm{RB}}$is the bandwidth of one PRB. In a practical implementation, this expression is replaced or calibrated by the CQI-to-MCS mapping and the corresponding transport block size.

This radio model is important for load balancing because a CIO adjustment may move a UE from a strong but congested serving cell to a less congested neighboring cell. Such a decision may reduce SINR slightly but still improve QoS if the target cell has more available resources.

*C. CIO-Biased Cell Association and Handover Model*

Cell Individual Offset is a configurable bias used in cell-selection and handover decisions. A positive CIO increases the effective attractiveness of a cell, while a negative CIO decreases it. Therefore, CIO values can be used to shift cell boundaries and redistribute users across neighboring cells.
Let

$$\mathrm{c}_t = [c_1(t), c_2(t), \dots, c_M(t)] \qquad (11)$$

denote the CIO vector at decision epoch $t$, where $c_i(t)$ is the CIO assigned to base station or sector $i$. Each CIO value is bounded by operational limits:

$$c_i(t) \in [c_{\min}, c_{\max}], i = 1, \dots, M. \qquad (12)$$

The CIO-biased association metric between UE $u$and base station $i$is defined as

$$\Phi_{u,i}(t) = Q_{u,i}(t) + c_i(t), \qquad (13)$$

where $Q_{u,i}(t)$ is a radio-quality indicator such as RSRP, RSRQ, SINR, or a CQI-derived metric. The CIO term modifies the effective cell-selection metric without changing the physical channel. As a result, the controller can influence handover behavior while still relying on standard radio measurements.
For a UE currently served by base station $s$, a neighboring base station $j$becomes a handover candidate if its biased metric exceeds that of the serving cell by a hysteresis margin:

$$\Phi_{u,j}(t) > \Phi_{u,s}(t) + H_{\mathrm{ys}}, \qquad (14)$$

where $H_{\mathrm{ys}}$ is the handover hysteresis parameter. To avoid handovers caused by short-term fading, the condition must remain satisfied for a Time-To-Trigger duration $T_{\mathrm{TTT}}$. If the condition persists for the required duration, the UE is handed over from the serving base station $s$to the target base station $j$.
This handover model captures an important trade-off. Increasing a cell's CIO can attract more users, which may improve load balance if the cell is underutilized. However, if CIO values are changed too aggressively, UEs near cell boundaries may repeatedly switch between neighboring cells, causing ping-pong handovers and degraded QoS. Therefore, CIO control must be both load-aware and stability-aware.

*D. Traffic, Queueing, and Scheduling Model*

Each UE has a downlink traffic queue at its serving base station. Let $q_u(t)$ denote the queue length for UE $u$ at time $t$. During each time interval, new packets arrive and the scheduler serves part of the queue depending on allocated resources and channel quality. The queue evolves as

$$q_u(t+1) = \max\{0, q_u(t) + A_u(t) - D_u(t)\}, \qquad (15)$$

where $A_u(t)$ is the amount of new traffic arriving for UE $u$, and $D_u(t)$ is the amount of data successfully served during the interval.
The traffic model may include full-buffer flows, bursty Poisson arrivals, or mixed service classes. Full-buffer traffic represents continuously active users with saturated queues, while bursty traffic captures more realistic applications such as web browsing, video segments, or intermittent data sessions. The MAC scheduler allocates PRBs to active UEs based on both channel quality and QoS urgency. A delay-aware proportional-fair scheduling score can be written as

$$\mathrm{score}_u(t) = \left(\frac{R_u(t)}{\bar{R}_u(t)}\right)^{1-\beta} \left(\frac{\mathrm{HOL}_u(t)}{\bar{\mathrm{HOL}}(t)}\right)^{\beta} w_{\mathrm{QCI}}(u), \qquad (16)$$

where $R_u(t)$ is the instantaneous achievable rate, $\bar{R}_u(t)$ is the exponentially averaged historical rate, $\mathrm{HOL}_u(t)$ is the head-of-line packet delay, $\bar{\mathrm{HOL}}(t)$ is the average HOL delay across active users, and $w_{\mathrm{QCI}}(u)$ is a service-class weight. The parameter $\beta$ controls the relative importance of channel-aware scheduling and delay-aware prioritization.
The scheduler determines the number of PRBs assigned to each UE, the served bits, packet delay, jitter, packet loss, and throughput. These metrics are then aggregated at the cell level and provided to the reinforcement learning agent as part of the observed state.
Traffic and scheduling are directly connected to load balancing. If too many UEs are associated with the same cell, resource contention increases, queues grow, latency rises, and packet loss becomes more likely. By adjusting CIO values, the controller can redistribute users and reduce congestion in overloaded cells.

*E. Gauss–Markov User Mobility Model*

User mobility plays a central role in load balancing and handover control. To generate realistic UE trajectories, we adopt the Gauss–Markov mobility model. This model produces temporally correlated motion, meaning that a UE's current speed and direction depend partly on its previous speed and direction. This is more realistic than a memoryless random-walk model, where users may abruptly change direction at every time step.
For UE $u$, let $v_u(t)$ denote its speed and $\theta_u(t)$ denote its movement direction at time $t$. The Gauss–Markov model updates speed and direction as

$$v_u(t) = \alpha v_u(t-1) + (1-\alpha)\bar{v} + \sqrt{1-\alpha^2}\,\delta_v(t), \qquad (17)$$

$$\theta_u(t) = \alpha\theta_u(t-1) + (1-\alpha)\bar{\theta} + \sqrt{1-\alpha^2}\delta_\theta(t), \quad (18)$$

where $\bar{v}$ is the average speed, $\bar{\theta}$ is the average movement direction, and

$$\delta_v(t) \sim \mathcal{N}(0,\sigma_v^2), \delta_\theta(t) \sim \mathcal{N}(0,\sigma_\theta^2), \quad (19)$$

The parameter $\alpha \in [0,1]$ controls the memory of the mobility process. When $\alpha = 0$, the movement is weakly correlated and changes more randomly. When $\alpha$ approaches one, the UE preserves its previous speed and direction more strongly, producing smoother trajectories. After updating speed and direction, the UE position is updated using

$$x_u(t) = x_u(t-1) + v_u(t)\Delta t\cos(\theta_u(t)), \quad (20)$$

$$y_u(t) = y_u(t-1) + v_u(t)\Delta t\sin(\theta_u(t)), \quad (21)$$

where $\Delta t$is the simulation time step. To avoid unrealistic movement, speeds are clipped to a feasible range, and heading angles are wrapped to a valid interval such as $(-\pi,\pi]$ or $[0,2\pi)$. Boundary handling is also applied to keep users inside the simulation region.

The Gauss–Markov model is particularly appropriate for evaluating load balancing because it creates realistic handover patterns. UEs may gradually approach cell boundaries, remain near overlap regions, or move along border areas for extended periods. These conditions are challenging for mobility management because small changes in radio measurements or CIO values may affect handover decisions. Therefore, the model allows us to evaluate whether the learned controller can reduce unnecessary handovers while still offloading users when congestion relief is beneficial.

### *F. Observation Model with Measurement Noise*

In a practical 5G radio access network, the controller does not observe the exact physical state of the system. Instead, it receives measurements that are noisy, quantized, delayed, and aggregated over reporting windows. This creates a mismatch between the true network state and the state available to the learning agent.

For example, RSRP measurements are influenced by receiver noise, interference, fading, and measurement filtering. CQI values are quantized indicators derived from SINR and are reported periodically. Latency and packet-loss measurements are computed from packet counters and queueing statistics, which may be delayed or averaged over time. If a reinforcement learning agent is trained only on ideal measurements, it may learn a policy that performs well in simulation but degrades under realistic reporting uncertainty.

To address this issue, we model the observation available to the agent as a corrupted and filtered version of the true network statistics. Let $x_t$ denote a generic per-cell aggregate at decision epoch $t$. This quantity may represent average RSRP, average CQI, average latency, jitter, packet-loss ratio, or cell utilization. The noisy observation is denoted by $\tilde{x}_t$.

For key radio and QoS metrics, the noisy measurements are modeled as

$$\widetilde{\mathrm{RSRP}}_t = \mathrm{RSRP}_t + \epsilon_t^{\mathrm{RSRP}}, \epsilon_t^{\mathrm{RSRP}} \sim \mathcal{N}(0,\sigma_{\mathrm{RSRP}}^2), \quad (22)$$

$$\widetilde{\mathrm{CQI}}_t = \mathrm{CQI}_t + \epsilon_t^{\mathrm{CQI}}, \epsilon_t^{\mathrm{CQI}} \sim \mathcal{N}(0,\sigma_{\mathrm{CQI}}^2), \quad (23)$$

$$\widetilde{D}_{\mathrm{avg},t} = D_{\mathrm{avg},t} + \epsilon_t^D, \epsilon_t^D \sim \mathcal{N}(0,\sigma_D^2). \quad (24)$$

The Gaussian noise model provides a controlled way to evaluate robustness under different levels of measurement uncertainty. Although real measurement errors may not be perfectly Gaussian, the additive Gaussian model is a practical approximation for aggregated reports, where multiple small error sources may combine into approximately zero-mean disturbances.

After noise injection, observations are projected back to their physical domains. CQI values are rounded and clipped to the valid CQI range. Delay, jitter, and packet-loss values are constrained to remain nonnegative. This prevents the learning agent from receiving physically impossible observations.

### *G. Temporal Filtering and Observation Normalization*

In addition to measurement noise, network reports are often delayed or smoothed because base stations compute statistics over finite time windows. A controller typically does not receive fully instantaneous measurements. To model this effect, we apply exponential moving average filtering to each noisy observation channel:

$$\hat{x}_t = \lambda\hat{x}_{t-1} + (1-\lambda)\tilde{x}_t, 0 \le \lambda < 1. \quad (25)$$

The parameter $\lambda$ controls the degree of smoothing. When $\lambda$ is close to zero, the controller observes measurements that are close to the latest noisy values. When $\lambda$ is larger, the observation becomes smoother but more delayed. This reflects the behavior of practical telemetry systems that report averaged statistics rather than instantaneous values.

This filtering is useful for mobility management because CIO decisions should not react to very short-lived fading spikes. Handover and load-balancing decisions typically operate on a slower time scale than physical-layer fluctuations. By using filtered observations, the controller is encouraged to respond to persistent trends such as sustained congestion, increasing delay, or stable neighboring-cell advantage.

The observation vector contains heterogeneous features with different units and magnitudes. For example, throughput may be measured in Mbps, delay in milliseconds, packet-loss ratio in percentages, and fairness as a unitless index. Without normalization, large-magnitude features may dominate neural-network training. Therefore, each filtered observation channel is normalized as

$$z_t = \frac{\hat{x}_t - \mu_t}{\sqrt{s_t^2 + \varepsilon}}, \quad (26)$$

where $\mu_t$ and $s_t^2$ are running estimates of the mean and variance, and $\varepsilon$ is a small numerical constant. During evaluation, the normalization statistics obtained during training are kept fixed. This ensures that the deployed policy receives inputs with the same scaling used during training.

### *H. Markov Decision Process Formulation*

The QoS-aware load-balancing problem is formulated as an episodic Markov Decision Process:

$$\mathcal{M} = (\mathcal{S}, \mathcal{A}, \mathcal{P}, \mathcal{R}, \gamma), \quad (27)$$

where $\mathcal{S}$ is the state space, $\mathcal{A}$ is the action space, $\mathcal{P}$ is the transition probability function, $\mathcal{R}$ is the reward function, and $\gamma$ is the discount factor.

At each decision epoch $t$, the controller observes the current state $\mathrm{s}_t$, selects an action $\mathrm{a}_t$, receives a reward $r_t$, and then observes the next state $\mathrm{s}_{t+1}$. The transition from $\mathrm{s}_t$ to $\mathrm{s}_{t+1}$ is determined by UE mobility, wireless propagation, traffic arrivals, scheduling, handover logic, and observation uncertainty.

The complete physical network state would include UE positions, velocities, channel gains, serving-cell assignments, queues, traffic arrivals, fading states, and previous handover timers. In practice, the controller does not observe all these variables directly. Instead, it receives aggregated and noisy measurements. Therefore, the learning problem is partially observable. In this work, the filtered and normalized measurement vector is treated as the effective MDP state for policy learning.

#### *1) State Space*

The state is constructed from per-cell aggregate statistics. Let $\eta_i(t)$ denote the PRB utilization of base station $i$, defined as the fraction of PRBs used during the latest measurement window. The utilization vector is

$$\eta(t) = [\eta_1(t), \eta_2(t), \dots, \eta_M(t)]. \quad (28)$$

Let $\bar{T}_i(t)$ denote the average downlink throughput served by base station $i$. The throughput vector is

$$\mathrm{T}(t) = [\bar{T}_1(t), \bar{T}_2(t), \dots, \bar{T}_M(t)]. \quad (29)$$

Let $\bar{L}_i(t)$, $\bar{J}_i(t)$, and $\mathrm{P\bar{L}R}_i(t)$ denote the average latency, average jitter, and packet-loss ratio for traffic served by base station $i$, respectively. These vectors are defined as

$$\mathrm{L}(t) = [\bar{L}_1(t), \bar{L}_2(t), \dots, \bar{L}_M(t)], \quad (30)$$

$$\mathrm{J}(t) = [\bar{J}_1(t), \bar{J}_2(t), \dots, \bar{J}_M(t)], \quad (31)$$

$$\mathrm{P}(t) = [\mathrm{P\bar{L}R}_1(t), \mathrm{P\bar{L}R}_2(t), \dots, \mathrm{P\bar{L}R}_M(t)]. \quad (32)$$

Recent handover activity is represented by

$$\mathrm{H}(t) = [\bar{H}_1(t), \bar{H}_2(t), \dots, \bar{H}_M(t)], \quad (33)$$

where $\bar{H}_i(t)$ indicates the recent number or rate of handovers associated with base station $i$. After noise injection, temporal filtering, and normalization, these vectors are concatenated to form the state:

$$\mathrm{s}_t = [\eta(t), \mathrm{T}(t), \mathrm{L}(t), \mathrm{J}(t), \mathrm{P}(t), \mathrm{H}(t)] \in \mathbb{R}^{6M}. \quad (34)$$

This state representation allows the agent to observe both radio-access load and user-experienced QoS. It enables the policy to learn, for example, whether a high-utilization cell is also experiencing high latency and packet loss, or whether a neighboring cell has enough capacity to accept additional users.

#### *2) Action Space*

The controller action is defined as a continuous CIO adjustment vector:

$$\mathrm{a}_t = \Delta\mathrm{c}_t = [\Delta c_1(t), \Delta c_2(t), \dots, \Delta c_M(t)]. \quad (35)$$

Each component is bounded by

$$\Delta c_i(t) \in [-\Delta c_{\max}, \Delta c_{\max}], i = 1, \dots, M. \quad (36)$$

The applied CIO vector is updated according to

$$\mathrm{c}_t = \mathrm{clip}(\mathrm{c}_{t-1} + \mathrm{a}_t, c_{\min}, c_{\max}). \quad (37)$$

This formulation has two advantages. First, it ensures that CIO values remain within feasible operational limits. Second, it prevents the controller from making unrealistically large changes in a single control interval. Since handover decisions are sensitive to bias values, gradual CIO changes are preferred for stable mobility management.

*3) Reward Function*

Each reward component is normalized to the interval $[0, 1]$. For benefit metrics, including throughput and Jain's fairness index, larger normalized values indicate better performance. For cost metrics, including latency, jitter, packet-loss ratio, and handover activity, the normalized reward component is inverted so that a larger value always indicates better performance. Therefore, $R_{LAT}$, $R_{JIT}$, $R_{PLR}$, and $R_{HO}$represent low-latency, low-jitter, low-loss, and low-handover utilities, respectively. The reward function is designed to capture the multi-objective nature of QoS-aware load balancing. The agent should increase throughput and fairness while reducing latency, jitter, packet loss, and unnecessary handovers. The scalar reward is then written as

$$R_t = w_1 R_{THR}(t) + w_2 R_{JIT}(t) + w_3 R_{LAT}(t) + w_4 R_{JF}(t) + w_5 R_{PLR}(t) + w_6 R_{HO}(t). \quad (38)$$

The weights satisfy

$$\sum_k w_k = 1, w_k \geq 0. \quad (39)$$

The handover-related term discourages excessive handover activity. CIO smoothness is handled separately through bounded CIO updates and the smoothness regularization term used during PPO training. This smoothness penalty discourages abrupt changes in handover bias and reduces the likelihood of ping-pong behavior.

Fairness is measured using Jain's fairness index. For a set of $n$throughput values $x_1(t), x_2(t), \ldots, x_n(t)$, the fairness index is

$$J(t) = \frac{\left(\sum_{i=1}^{n} x_i(t)\right)^2}{n \sum_{i=1}^{n} x_i^2(t)}, J(t) \in \left[\frac{1}{n}, 1\right]. \quad (40)$$

A value close to one indicates balanced service, while a smaller value indicates that resources are unevenly distributed.

The reward therefore encourages the agent to find a balanced operating point. It should not maximize throughput by overloading only the best radio cells, and it should not reduce handovers by keeping users connected to congested cells. Instead, it learns to trade off efficiency, fairness, reliability, latency, and mobility stability.

*4) Transition Dynamics*

The transition probability $\mathcal{P}(s_{t+1} \mid s_t, a_t)$ is induced by the network simulator. Given the current state and action, the simulator evolves the system through the following sequence.

First, UE mobility updates user positions and velocities. Second, the propagation model updates path loss, shadowing, fading, received powers, SINR, CQI, and achievable rates. Third, traffic arrivals update the queues of active users. Fourth, the MAC scheduler allocates PRBs and serves packets based on channel quality and QoS urgency. Fifth, the handover module evaluates CIO-biased cell-selection metrics and updates serving-cell assignments when handover conditions are satisfied. Finally, the resulting QoS metrics are aggregated, corrupted by measurement noise, filtered, normalized, and returned as the next state.

The transition is stochastic because mobility perturbations, traffic arrivals, fading, shadowing, and measurement noise are random. The action affects the transition indirectly through CIO values. By changing CIO values, the agent changes the probability and timing of future handovers, which changes the distribution of UEs across cells. This affects cell load, PRB utilization, queueing delay, throughput, packet loss, and future handover behavior.

*5) Discount Factor and Learning Objective*

The goal is to learn a stationary policy $\pi(a \mid s)$ that maps each observed state to a distribution over feasible CIO adjustment actions. The optimal policy maximizes the expected discounted return:

$$\pi^* = \arg\max_{\pi} \mathbb{E}_{\pi}\left[\sum_{t=0}^{T-1} \gamma^t r_t\right], \quad (41)$$

where $T$ is the episode length and $\gamma \in (0,1)$ is the discount factor.

A high discount factor is appropriate because CIO actions often have delayed consequences. A CIO adjustment may not immediately change user association because handover hysteresis and TTT must be satisfied. Even after a handover occurs, the effect on queueing delay, packet loss, and fairness may appear over several future intervals. Therefore, the agent must optimize long-term QoS rather than reacting only to immediate radio measurements.

In this work, the learning objective encourages the controller to select CIO adjustments that produce sustained improvements in network performance. The learned policy is expected to reduce congestion, improve resource utilization, preserve fairness, reduce delay and packet loss, and avoid unnecessary handover instability under mobility and observation uncertainty

## IV. Proposed Method

This section presents the proposed Proximal Policy Optimization based framework for QoS-aware load balancing in 5G cellular networks. Based on the MDP formulation introduced in Section III, the objective is to learn a control policy that dynamically adjusts Cell Individual Offset (CIO) values in order to influence user association and handover behavior. The learned controller observes the current network state, selects a bounded CIO adjustment action, and receives a reward that reflects the resulting network-wide QoS performance.

The central idea is to use CIO as an indirect but practical control variable for load balancing. Instead of directly forcing each UE to connect to a specific base station, the controller modifies the handover bias of each cell. These CIO values shift the effective attractiveness of cells during handover decisions. As a result, UEs near cell boundaries can be gradually steered toward less congested neighboring cells when doing so improves the overall QoS. This makes the proposed method compatible with conventional handover logic while adding an intelligent optimization layer.

The proposed method is designed for realistic mobile network conditions. In a 5G RAN, the effect of a load-balancing decision is not immediate. A CIO adjustment may influence handover only after the A3 condition remains valid for the Time-To-Trigger duration. After the handover occurs, queueing delay, packet loss, and throughput may change over several future scheduling intervals. Therefore, the controller must optimize long-term QoS rather than reacting only to the immediate radio measurements. PPO is selected because it supports continuous control, provides stable policy updates through a clipped objective, and uses an actor–critic structure that can learn from delayed rewards.

The proposed framework consists of five main components: a bounded CIO adjustment policy, an actor–critic neural architecture, advantage estimation, clipped PPO optimization, and reward-stability mechanisms. These components are described in detail below.

### *A. PPO-Based Load-Balancing Framework*

At each decision epoch $t$, the controller receives the normalized network state $\mathrm{s}_t$. This state contains the per-cell statistics defined in Section III, including cell utilization, average throughput, latency, jitter, packet-loss ratio, and recent handover activity. Based on this state, the PPO actor generates a continuous action vector:

$$\mathrm{a}_t = \Delta\mathrm{c}_t = [\Delta c_1(t), \Delta c_2(t), \ldots, \Delta c_M(t)]. \qquad (42)$$

Here, $M$ is the number of controlled cells or sectors, and $\Delta c_i(t)$ is the CIO adjustment applied to cell $i$. The action is not the absolute CIO value; it is the incremental change applied to the previous CIO configuration. This design is important because it prevents abrupt changes in handover bias and encourages smooth load-balancing behavior. The CIO vector applied to the network is updated as

$$\mathrm{c}_t = \mathrm{clip}(\mathrm{c}_{t-1} + \Delta\mathrm{c}_t, c_{\min}, c_{\max}), \qquad (43)$$

where $\mathrm{c}_{t-1}$ is the previous CIO vector, and $c_{\min}$ and $c_{\max}$ are the minimum and maximum allowable CIO values. The clipping operation ensures that each CIO remains within the valid operational range.

After the CIO vector is updated, the new CIO values are used by the handover module. They modify the CIO-biased cell-selection metric and therefore influence future UE association. The environment then evolves over the next control interval: UEs move, channel conditions change, traffic arrives, PRBs are scheduled, queues evolve, and handovers may occur. At the end of the interval, the simulator computes the resulting QoS metrics and returns a scalar reward $r_t$. The transition $(\mathrm{s}_t, \Delta\mathrm{c}_t, r_t, \mathrm{s}_{t+1})$ is stored and later used to update the PPO policy.

This closed-loop learning process allows the agent to discover how CIO adjustments affect network behavior. For example, if one cell is heavily loaded and has increasing latency, the policy may learn to reduce that cell's CIO. This makes the cell less attractive to UEs near the cell boundary and can gradually shift users toward neighboring cells. Similarly, if another cell has available PRBs and acceptable radio quality, the policy may increase its CIO to attract more users. Through training, the policy learns how to balance these decisions across all cells.

### *B. Motivation for Using PPO*

The load-balancing problem considered in this work has several properties that make PPO suitable.

First, the action space is continuous. CIO values are usually expressed in dB and can be adjusted over a range of values. Discretizing CIO actions may limit control resolution and can become inefficient when multiple cells are controlled jointly. For $M$ cells, even a small number of discrete CIO levels can produce a large combinatorial action space. PPO avoids this limitation by directly learning a continuous policy.

Second, the environment is stochastic. The same CIO adjustment can produce different outcomes depending on UE mobility, traffic arrivals, fading, interference, scheduling decisions, and observation noise. A stochastic actor policy is therefore useful during training because it allows the controller to explore different bias configurations and avoid premature convergence.

Third, the effect of an action is delayed. CIO changes do not instantly change QoS. They affect handover probability, UE association, resource sharing, queue evolution, and packet delivery over future intervals. PPO uses a critic and advantage estimation to learn from these delayed effects.

Fourth, stability is essential. In mobility management, aggressive policy changes can cause unstable association boundaries and ping-pong handovers. PPO uses a clipped surrogate objective that limits how much the policy can change during each update. This is especially useful when the reward signal is noisy or when observation uncertainty exists.

For these reasons, PPO provides a suitable balance between continuous control capability, exploration, learning stability, and robustness.

*C. Continuous CIO Adjustment Policy*

The actor network represents a stochastic policy $\pi_\theta(\Delta \mathrm{c}_t \mid \mathrm{s}_t)$ where $\theta$ denotes the actor parameters. The policy maps the current network state $\mathrm{s}_t$ to a probability distribution over CIO adjustment actions.

Since the action has $M$ continuous components, the policy is modeled as a diagonal Gaussian distribution in an unconstrained latent space. Given the input state $\mathrm{s}_t$, the actor outputs a mean vector

$$\mu_\theta(\mathrm{s}_t) = [\mu_1(\mathrm{s}_t), \mu_2(\mathrm{s}_t), \dots, \mu_M(\mathrm{s}_t)] \qquad (44)$$

and a standard deviation vector

$$\sigma_\theta(\mathrm{s}_t) = [\sigma_1(\mathrm{s}_t), \sigma_2(\mathrm{s}_t), \dots, \sigma_M(\mathrm{s}_t)]. \qquad (45)$$

A latent action vector is sampled as

$$\mathrm{u}_t = \mu_\theta(\mathrm{s}_t) + \sigma_\theta(\mathrm{s}_t) \odot \epsilon_t, \epsilon_t \sim \mathcal{N}(0, \mathrm{I}), \qquad (46)$$

where $\odot$ denotes element-wise multiplication. This sampling process allows the policy to explore different CIO adjustments during training.

Because $\mathrm{u}_t$ is unbounded, it cannot be directly applied to the network. A large unbounded action could generate an unrealistic CIO change and destabilize the handover process. Therefore, the latent action is passed through a hyperbolic tangent function $\hat{\mathrm{a}}_t = \tanh(\mathrm{u}_t)$. The output $\hat{\mathrm{a}}_t$ lies in $(-1,1)^M$. It is then scaled to the maximum allowable CIO adjustment per control interval:

$$\Delta \mathrm{c}_t = \Delta c_{\max} \hat{\mathrm{a}}_t. \qquad (47)$$

Thus, each action component satisfies

$$\Delta c_i(t) \in [-\Delta c_{\max}, \Delta c_{\max}]. \qquad (48)$$

This bounded-action formulation has an important practical interpretation. The parameter $\Delta c_{\max}$defines how much the controller is allowed to change each cell's CIO within one control interval. A small value of $\Delta c_{\max}$enforces conservative and smooth adaptation, while a larger value allows faster but potentially less stable load redistribution. In this work, the bounded adjustment structure is used to ensure that the controller changes CIO values gradually and avoids sudden handover-boundary shifts.

*D. Actor–Critic Neural Architecture*

The PPO controller uses an actor–critic architecture. The actor learns the CIO adjustment policy, while the critic estimates the expected long-term return from a given state. The critic is represented by a value function $V_\psi(\mathrm{s}_t)$ where $\psi$ denotes the critic parameters. The critic approximates

$$V_\psi(\mathrm{s}_t) \approx \mathbb{E}_\pi\left[ \sum_{k=0}^{\infty} \gamma^k r_{t+k} \,\middle|\, \mathrm{s}_t \right], \qquad (49)$$

where $\gamma \in (0,1)$ is the discount factor. The value function estimates how good the current network state is under the current policy.

The critic is needed because the reward depends on future consequences of current CIO actions. For example, increasing the CIO of a lightly loaded cell may not immediately improve throughput. However, after some UEs hand over to that cell, the overloaded neighboring cell may experience lower queueing delay and lower packet loss. The critic helps the agent learn such delayed relationships by estimating the long-term return.

Both the actor and critic receive the same normalized state vector as input. The input includes the filtered and normalized per-cell features:

$$\mathrm{s}_t = [\eta(t), \mathrm{T}(t), \mathrm{L}(t), \mathrm{J}(t), \mathrm{P}(t), \mathrm{H}(t)]. \qquad (50)$$

In the default implementation, both networks are implemented as feedforward multilayer perceptrons. Each network contains two hidden layers with 256 neurons per layer and ReLU activation functions. The actor output layer produces the Gaussian policy parameters, while the critic output layer produces a single scalar value.

The actor outputs the mean of the Gaussian distribution and a log-standard-deviation parameter. The standard deviation is kept positive by exponentiating or otherwise transforming the log-standard deviation. In practice, the log-standard-deviation can be bounded to avoid numerical instability. If the standard deviation becomes too small too early, the policy may stop exploring. If it becomes too large, the action samples may become noisy and unstable. Bounding this parameter helps maintain useful exploration during training.

This neural architecture is intentionally lightweight. Once training is complete, the online controller only needs to run a forward pass through the actor network to compute CIO adjustments. Therefore, the trained policy can be executed efficiently at each control interval.

*E. Advantage Estimation*

PPO updates the policy using advantage estimates. The advantage function measures whether the selected action performed better or worse than the expected behavior in the same state. It is defined as

$$A^{\pi}(\mathrm{s}_t, \Delta \mathrm{c}_t) = Q^{\pi}(\mathrm{s}_t, \Delta \mathrm{c}_t) - V^{\pi}(\mathrm{s}_t). \quad (52)$$

A positive advantage means that the selected CIO adjustment produced a better outcome than expected. A negative advantage means that the action produced a worse outcome.

Directly estimating the action-value function $Q^{\pi}$ is difficult in this problem. The environment includes random mobility, fading, traffic arrivals, queueing, scheduling, observation noise, and delayed handover effects. Therefore, the reward following one action can be noisy, and the full effect of the action may appear only after several future intervals.

To obtain stable advantage estimates, we use Generalized Advantage Estimation (GAE). First, the one-step temporal-difference residual is computed as

$$\delta_t = r_t + \gamma V_{\psi}(\mathrm{s}_{t+1}) - V_{\psi}(\mathrm{s}_t). \quad (53)$$

The term $\delta_t$ measures the difference between the observed one-step outcome and the critic's prediction. If $\delta_t$ is positive, the transition was better than predicted. If it is negative, the transition was worse than predicted.

GAE then computes the advantage estimate by accumulating future TD residuals:

$$\hat{A}_t = \sum_{l=0}^{T-t-1} (\gamma\lambda)^l \delta_{t+l}, \quad (53)$$

where $T$ is the rollout horizon and $\lambda \in [0,1]$ controls the bias–variance trade-off. When $\lambda$ is small, the estimator relies mostly on short-term TD errors, which reduces variance but may increase bias. When $\lambda$ is close to one, the estimator includes longer-term information, which reduces bias but may increase variance.

In the considered load-balancing problem, delayed effects are important. CIO adjustments may influence handovers, queueing delay, and packet loss over multiple future intervals. Therefore, a relatively large value such as $\lambda = 0.95$ is suitable because it allows the advantage estimate to capture medium-term QoS effects while maintaining stable learning. The target value used to train the critic is

$$\hat{V}_t^{\mathrm{tgt}} = \hat{A}_t + V_{\psi}(\mathrm{s}_t). \quad (54)$$

This target combines the critic's current estimate with the multi-step advantage correction. The critic is trained to reduce the error between $V_{\psi}(\mathrm{s}_t)$ and $\hat{V}_t^{\mathrm{tgt}}$.

### *F. PPO Clipped Surrogate Objective*

The main policy update in PPO is based on the clipped surrogate objective. Let $\theta_{\mathrm{old}}$ denote the policy parameters used to collect the rollout data. After collecting a rollout, the policy is updated to new parameters $\theta$. The probability ratio between the new and old policies is

$$\rho_t(\theta) = \frac{\pi_{\theta}(\Delta \mathrm{c}_t \mid \mathrm{s}_t)}{\pi_{\theta_{\mathrm{old}}}(\Delta \mathrm{c}_t \mid \mathrm{s}_t)}. \quad (55)$$

This ratio measures how much the probability of the sampled action changes after the policy update. If $\rho_t(\theta) > 1$, the new policy gives higher probability to the action than the old policy. If $\rho_t(\theta) < 1$, the new policy gives lower probability to the action.

The clipped PPO objective is

$$L^{\mathrm{clip}}(\theta) = \mathbb{E}_t\left[\min\left(\rho_t(\theta)\hat{A}_t, \mathrm{clip}(\rho_t(\theta), 1-\epsilon, 1+\epsilon)\hat{A}_t\right)\right], \quad (56)$$

where $\epsilon$ is the clipping parameter.

The clipping operation prevents the policy from changing too much in one optimization step. This is important because the rollout data were collected using the old policy. If the new policy becomes too different from the old policy, the collected data may no longer provide reliable learning information.

In the context of CIO control, clipped updates also have a useful control interpretation. The policy should not suddenly change from one handover strategy to a completely different one. Such instability could cause large variations in CIO adjustments and increase handover oscillations. The PPO clipping mechanism encourages gradual policy improvement, which is aligned with the need for stable mobility management.

### *G. Entropy Bonus and Critic Loss*

To maintain exploration, PPO includes an entropy bonus. The entropy term is defined as

$$L^{\mathrm{ent}}(\theta) = \mathbb{E}_t[\mathcal{H}(\pi_{\theta}(\cdot \mid \mathrm{s}_t))], \quad (57)$$

where $\mathcal{H}(\cdot)$ denotes the entropy of the policy distribution. Higher entropy means that the policy is more exploratory. Lower entropy means that the policy is more deterministic.

Entropy regularization is useful during training because the agent initially does not know which CIO patterns are effective. It must explore different strategies, such as offloading users earlier, delaying handovers, increasing the attractiveness of lightly loaded cells, or reducing the attractiveness of congested cells. Without sufficient exploration, the policy may converge too early to a suboptimal strategy.

The critic is trained using the value loss

$$L^V(\psi) = \mathbb{E}_t\left[\left(V_\psi(\mathrm{s}_t) - \hat{V}_t^{\mathrm{tgt}}\right)^2\right]. \tag{58}$$

The value loss encourages the critic to accurately predict the expected return. A more accurate critic produces better advantage estimates, which improves policy-gradient learning. The final PPO objective is written as

$$\max_{\theta,\psi} \quad L^{\mathrm{clip}}(\theta) + c_{\mathrm{ent}} L^{\mathrm{ent}}(\theta) - c_v L^V(\psi), \tag{59}$$

where $c_{\mathrm{ent}}$ is the entropy coefficient and $c_v$ is the value-loss coefficient. The actor parameters are updated to maximize the policy objective, while the critic parameters are updated to minimize the value prediction error.

### *H. Reward Processing and Learning Stability*

The reward used by the PPO agent is the multi-objective QoS reward defined in Section III. It combines throughput, fairness, latency, jitter, packet-loss ratio, and handover rate. Since these metrics have different units and different desirable directions, reward processing is required before training.

Throughput and fairness are higher-is-better metrics. Latency, jitter, packet-loss ratio, and handover rate are lower-is-better metrics. Therefore, all KPI values are first mapped to normalized utilities in $[0, 1]$, where larger values always indicate better performance. This makes the reward components comparable and prevents one metric from dominating the learning signal only because of its numerical scale. The PPO agent receives the scalar reward as (38). In addition to the scalar QoS reward in Eq. (38), a separate CIO smoothness regularization term is applied during PPO training to discourage large CIO changes in a single control interval. It discourages the policy from making large CIO changes in a single interval. This is necessary because aggressive CIO changes can shift cell boundaries too strongly and may cause ping-pong handovers.

The smoothness penalty also helps the policy learn a more practical control behavior. Instead of reacting strongly to every short-term measurement fluctuation, the agent is encouraged to make smaller and more consistent adjustments. This is especially important under observation noise, where temporary measurement errors could otherwise cause unnecessary CIO changes.

Some QoS metrics, especially latency, jitter, and packet loss, may contain extreme spikes during congestion or handover instability. If these extreme values are used directly, they may produce very large advantage estimates and destabilize learning. To avoid this, KPI utilities are clipped using reference bounds or robust percentiles. This preserves sensitivity in the normal operating range while limiting the effect of rare outliers.

Handover-related penalties are also treated carefully. A single handover is not always bad; it may be necessary to relieve congestion or improve signal quality. However, repeated handovers in a short time indicate instability. Therefore, the handover utility can be computed over a recent window or smoothed using an exponential moving average. This allows the reward to penalize persistent handover churn rather than isolated useful handovers.

### *I. Training Procedure*

The PPO agent is trained through repeated interaction with the network simulator. Each training iteration consists of rollout collection, advantage estimation, and policy optimization.

At the beginning of each episode, the simulator initializes UE positions, velocities, serving-cell assignments, traffic queues, channel conditions, and CIO values. The initial CIO vector may be set to zero or to a predefined baseline configuration.

At each decision epoch $t$, the simulator constructs the normalized state vector $\mathrm{s}_t$. The actor network samples a CIO adjustment action $\Delta\mathrm{c}_t$. The CIO vector is updated using

$$\mathrm{c}_t = \mathrm{clip}(\mathrm{c}_{t-1} + \Delta\mathrm{c}_t, c_{\min}, c_{\max}). \tag{60}$$

The updated CIO vector is applied to the handover logic. During the next control interval, UEs move according to the mobility model, channel conditions are updated, packets arrive, PRBs are allocated, queues evolve, and handovers are executed if their triggering conditions are satisfied. At the end of the interval, the environment computes the next state and reward.

The transition sample is stored in the rollout buffer. After collecting a fixed number of steps, GAE is used to compute advantage estimates. The rollout data are then divided into mini-batches, and the actor and critic are updated for multiple optimization epochs.

During optimization, the clipped PPO objective restricts policy changes. Gradient clipping is also applied to prevent numerical instability. In addition, the approximate KL divergence between the old and new policies can be monitored. If the KL divergence exceeds a predefined threshold, the update is stopped early to prevent excessive policy drift.

Training continues until the cumulative reward and the main QoS metrics converge. After training, the actor policy is frozen and evaluated on independent scenarios. During evaluation, the policy does not learn; it only maps the observed state to CIO adjustment actions.

### *J. Interpretation of the Learned CIO Policy*

After training, the learned policy acts as an adaptive CIO controller. Its decisions can be interpreted through the network conditions observed in the state.

When a cell has high PRB utilization, increasing latency, or increasing packet loss, the policy may reduce its CIO. This makes the cell less attractive in the handover process and can encourage UEs near the boundary to move toward neighboring cells. This reduces congestion and improves queueing performance.

When a cell has low utilization and acceptable radio quality, the policy may increase its CIO. This makes the cell more attractive and allows it to accept additional users. As a result, the network can use available resources more efficiently and improve fairness. When the network is already balanced, the policy may output small CIO adjustments close to zero. This prevents unnecessary handover-boundary movement and helps maintain stable associations.

The policy therefore does not simply maximize signal strength. Instead, it learns a trade-off between radio quality and network load. It may prefer a slightly weaker cell if that cell provides better resource availability and lower delay. This behavior is desirable in dense 5G networks, where strongest-signal association can create overloaded cells and poor QoS.

### *K. Expected Benefits of the Proposed Method*

The proposed PPO-based controller provides several benefits for QoS-aware 5G load balancing. First, it enables continuous CIO control. This allows fine-grained adjustment of handover biases and avoids the limitations of coarse discrete action spaces.

Second, it optimizes long-term QoS. By using a discounted return and advantage estimation, the controller learns the delayed effects of CIO actions on handovers, queues, packet loss, and fairness.

Third, it improves stability. PPO's clipped objective prevents overly large policy updates, while the smoothness penalty discourages aggressive CIO changes. Together, these mechanisms reduce the likelihood of ping-pong handovers and oscillatory control.

Fourth, it supports multi-objective optimization. The reward combines throughput, fairness, latency, jitter, packet loss, and handover behavior. This allows the policy to balance competing network objectives rather than optimizing a single metric.

Fifth, it is robust to noisy observations. The policy is trained using filtered and noisy network measurements, making it less dependent on perfect state information.

Overall, the proposed method learns a self-optimizing CIO adjustment policy that continuously adapts to user mobility, traffic variation, and changing radio conditions. By combining continuous bounded actions, actor–critic learning, GAE, clipped PPO updates, reward normalization, and CIO smoothness regularization, the proposed framework provides a stable and adaptive approach for dynamic load balancing in 5G cellular networks. Figure 1 illustrates the overall closed-loop structure of the proposed PPO-based QoS-aware CIO control framework. The 5G RAN environment includes user mobility, channel and interference variation, traffic queues, QoS-aware MAC scheduling, A3-based handover logic, and observation uncertainty. At each decision epoch, per-cell network measurements such as utilization, throughput, latency, jitter, packet-loss ratio, and recent handover activity are collected, filtered, and normalized to form the state vector provided to the PPO agent.

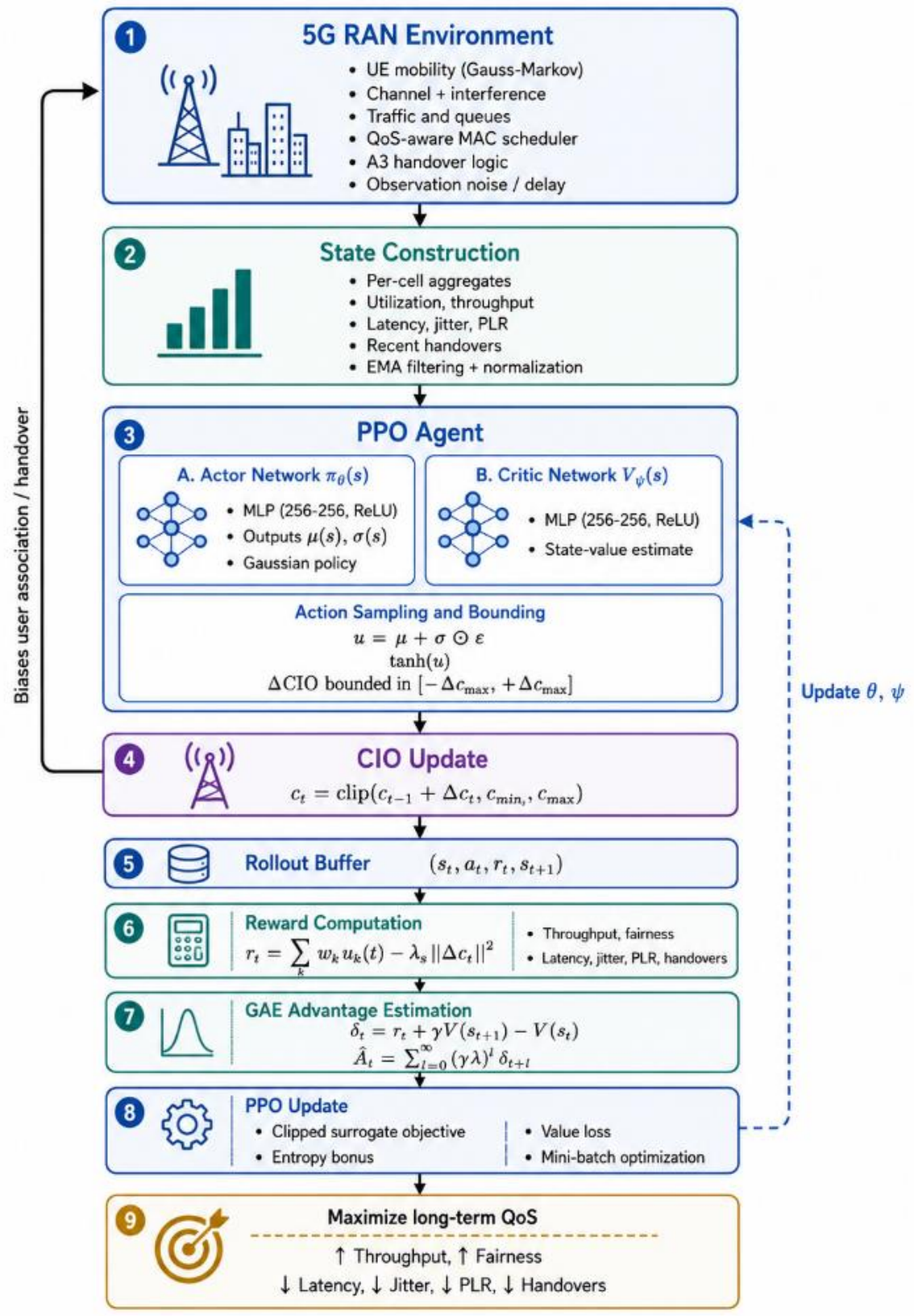

**Figure 1.** PPO-based QoS-aware CIO control framework for adaptive 5G load balancing. [71]

The PPO agent consists of an actor network and a critic network. The actor generates a stochastic continuous action representing the CIO adjustment vector, while the critic estimates the state value used for advantage estimation. The sampled action is bounded using a squashing function and then applied to update the CIO vector within operational limits. The updated CIO values bias user association and handover decisions in the RAN environment, thereby influencing future load distribution and QoS performance.

During training, the collected transitions are stored in a rollout buffer and used to compute the multi-objective reward, which combines throughput, fairness, latency, jitter, packet loss, handover activity, and CIO smoothness. Generalized Advantage Estimation (GAE) is then used to compute advantage values, and the PPO update optimizes the actor and critic using the clipped surrogate objective, entropy regularization, and value loss. This loop enables the controller to learn stable CIO adjustment policies that improve long-term

QoS while reducing unnecessary handovers and oscillatory control behavior.

## V. Performance Evaluation

This section evaluates the proposed PPO-based QoS-aware load-balancing framework under mobility, traffic variation, channel uncertainty, and noisy observations. The evaluation is designed to answer four main questions:

1. whether PPO can learn a stable CIO control policy;
2. whether the learned policy improves QoS compared with classical and learning-based baselines;
3. how the policy behaves under increasing user density;
4. how the learned policy behaves under the considered noisy-observation, mobility, scalability, and reward-ablation settings.

Unless otherwise stated, all algorithms are evaluated under the same topology, traffic model, mobility traces, channel conditions, observation-noise settings, and KPI definitions. This ensures that performance differences are caused by the control policies rather than by different simulation conditions.

### *A. Reward Normalization and Weight Selection*

The PPO agent is trained using a multi-objective QoS reward that combines six network performance indicators: throughput, fairness, latency, jitter, packet-loss ratio, and handover activity. These metrics have different units, numerical ranges, and optimization directions. For example, throughput and fairness should be maximized, while latency, jitter, packet loss, and handovers should be minimized. Therefore, before scalarizing the reward, each KPI is converted into a normalized utility value in the interval [0,1], where a larger value always represents better performance.

Let $\mathcal{K} = \{\text{thr,fair,lat,jit,plr,ho}\}$ denote the set of reward components. For each KPI $k \in \mathcal{K}$, the raw value at decision epoch $t$ is denoted by $x_{k,t}$. The normalized utility $\hat{x}_{k,t}$ is computed as

$$\hat{x}_{k,t} = \begin{cases} \text{clip}\left(\dfrac{x_{k,t} - m_k}{M_k - m_k}, 0, 1\right), & \text{if higher is better,} \\ \text{clip}\left(\dfrac{M_k - x_{k,t}}{M_k - m_k}, 0, 1\right), & \text{if lower is better.} \end{cases} \tag{61}$$

Here, $m_k$ and $M_k$ are the 5th and 95th percentile reference values of KPI $k$, computed from a held-out validation set.

### *B. Reproducible Reward-Weight Calibration*

To avoid hand-picking the reward weights, we use a reproducible multi-stage selection procedure. The goal is to choose a reward-weight vector that provides a balanced QoS trade-off rather than optimizing one KPI at the expense of others.

Let $\mathrm{w} = [w_{\text{thr}}, w_{\text{fair}}, w_{\text{lat}}, w_{\text{plr}}, w_{\text{jit}}, w_{\text{ho}}]$ denote a candidate reward-weight vector. For each candidate w, PPO is trained using the same training budget and evaluated over multiple independent random seeds. The evaluation produces a KPI summary vector $\mathrm{m}(\mathrm{w})$, which contains the mean performance values of throughput, fairness, latency, packet-loss ratio, jitter, and handover activity over the evaluation horizon.

Because the KPIs have different units and directions, each KPI summary is first mapped into a normalized utility value $\tilde{u}_k(\mathrm{w}) \in [0,1]$, where larger values always indicate better performance. For beneficial metrics such as throughput and fairness, the normalized utility is

$$\tilde{u}_k(\mathrm{w}) = \text{clip}\left(\frac{x_k(\mathrm{w}) - x_k^{\min}}{x_k^{\max} - x_k^{\min}}, 0, 1\right), \tag{62}$$

For cost metrics such as latency, jitter, packet-loss ratio, and handovers, the utility is inverted:

$$\tilde{u}_k(\mathrm{w}) = 1 - \text{clip}\left(\frac{x_k(\mathrm{w}) - x_k^{\min}}{x_k^{\max} - x_k^{\min}}, 0, 1\right). \tag{63}$$

This transformation makes all KPI utilities comparable and ensures that the weight vector represents an actual design preference rather than compensating for unit differences.

The candidate weights are selected from a coarse simplex grid subject to $\sum_k w_k = 1, w_k \geq 0$. For each candidate weight vector, PPO is trained and then evaluated using identical validation conditions. After evaluation, we apply a feasibility filter based on QoS constraints. A candidate is considered feasible if

$$\text{Latency}(\mathrm{w}) \leq 25 \text{ ms}, \text{PLR}(\mathrm{w}) \leq 5\%, \tag{64}$$

These constraints are applied to the mean evaluation metrics over the validation horizon. Tail-latency behavior is reported separately in the robustness analysis. The feasibility filter prevents selecting weight combinations that artificially increase throughput while tolerating unacceptable queueing delay or packet loss.

Among the feasible candidates, we then apply a Pareto-screening step. A candidate $\mathrm{w}_1$ is said to dominate another candidate $\mathrm{w}_2$ if $\tilde{u}_k(\mathrm{w}_1) \geq \tilde{u}_k(\mathrm{w}_2) \forall k$, and $\tilde{u}_j(\mathrm{w}_1) > \tilde{u}_j(\mathrm{w}_2)$ for at least one KPI $j$. Only non-dominated candidates are retained. This avoids choosing a reward configuration that is strictly worse than another feasible configuration across the KPI set.

Finally, a single operating point is selected from the non-dominated set by minimizing the distance to the ideal utility vector:

$$\mathrm{w}^\star = \arg\min_{\mathrm{w} \in \mathcal{P}} \| \tilde{\mathrm{u}}^{\text{ideal}} - \tilde{\mathrm{u}}(\mathrm{w}) \|_2, \tag{65}$$

where $\mathcal{P}$ is the Pareto set, and $\tilde{u}^{\text{ideal}}$ is the component-wise best utility observed among feasible candidates:

$$\tilde{u}_k^{\text{ideal}} = \max_{w \in \mathcal{F}} \tilde{u}_k(w). \qquad (66)$$

Here, $\mathcal{F}$denotes the feasible candidate set. This selection rule chooses a balanced operating point that remains close to the best achievable utility values across all KPIs, without over-optimizing a single metric. The resulting reward weights are shown in Table 1.

Table 1. The final weights

| $w_{\text{thr}}$ | $w_{\text{fair}}$ | $w_{\text{lat}}$ | $w_{\text{plr}}$ | $w_{\text{jit}}$ | $w_{\text{ho}}$ |
|---|---|---|---|---|---|
| $0.35$ | 0.20 | 0.20 | 0.1 | 0.1 | 0.05 |

The selected weights reflect the intended operating behavior of the controller. Throughput receives the largest weight because capacity improvement is a primary goal of load balancing. Latency and fairness are also strongly weighted because congestion control and balanced service are essential for user-perceived QoS. Packet-loss ratio and jitter receive moderate weights, as they are strongly affected by congestion and queueing instability. The handover weight is intentionally smaller because handover stability is also controlled through the CIO smoothness penalty. This prevents the agent from becoming overly conservative; excessive handover penalization may keep UEs attached to overloaded cells even when a controlled handover would improve system-level QoS.

To evaluate sensitivity to reward design, we also conduct controlled ablation experiments. These include uniform weights, removal of the fairness term, doubling the handover penalty, removing the CIO smoothness penalty, and perturbing individual reward weights by $\pm 50\%$. After each perturbation, the weights are renormalized to satisfy $\sum_k w_k = 1$. These experiments assess whether the proposed PPO gains are robust to reasonable changes in the reward specification.

*C. Fairness Metric*

Fairness is measured using Jain's fairness index. For a set of $n$served-rate values $x_1, x_2, \dots, x_n$, the fairness index is

$$J = \frac{\left(\sum_{i=1}^{n} x_i\right)^2}{n \sum_{i=1}^{n} x_i^2}, J \in \left[\frac{1}{n}, 1\right], \qquad (67)$$

In the default cell-level evaluation, $x_i$represents the cell-aggregate downlink throughput of cell $i$, averaged over the evaluation window. A value close to one indicates balanced service across cells, while a lower value indicates load imbalance or throughput concentration in a subset of cells. In additional user-level analysis, the same index may be computed over per-UE throughput values to quantify fairness among users.

*D. Robustness Stress-Test Design*

To examine the sensitivity of the proposed controller, we define a set of stress-test conditions involving observation quality, reporting behavior, and mobility variation. These tests are intended to evaluate how the learned policy behaves when the observed state becomes noisy, delayed, or partially unavailable. Unless explicitly reported in the result tables or figures, these stress-test settings should be interpreted as part of the evaluation design rather than as full deployment-level robustness evidence.

For measurement noise, we vary the standard deviation of the reported radio and QoS measurements as $\sigma_{\text{RSRP}} \in \{1,3,5\}$ dB, $\sigma_{\text{CQI}} \in \{0.5,1,2\}$ and $\sigma_{\text{lat/jit}} \in \{2,5,10\}$ ms. The noise is added independently to the reported measurements before filtering and normalization. This evaluates whether the policy can remain stable when telemetry is imperfect. For reporting delay, we introduce observation lag using a FIFO buffer with delays $L \in \{0,100,200,500\}$ ms. For missing observations, we randomly remove reported values with probability $p_{\text{miss}} \in \{0,1,5,10\}\%$. Missing values are imputed using zero-order hold, meaning that the most recently available observation is reused. This models practical telemetry systems in which some reports may be delayed or unavailable.

For mobility robustness, we evaluate three representative mobility regimes: pedestrian mobility at 1.5 m/s, urban vehicular mobility at 15m/s, and highway mobility at 30m/s. We also vary the Gauss–Markov memory parameter as $\alpha \in \{0.2,0.6,0.9\}$, to test different levels of trajectory smoothness. Lower $\alpha$values create more rapidly changing motion, while higher values produce smoother and more correlated trajectories.

For each perturbation setting, we report KPI degradation curves and confidence intervals. The evaluated KPIs include throughput, latency, jitter, packet-loss ratio, fairness, handover count, ping-pong rate, handover failure rate, and tail latency. Tail latency is reported using the 95th and 99th percentiles over the evaluation window.

*E. Generalization Considerations and Future Transfer Evaluation*

Generalization across unseen network layouts is an important requirement for practical RAN deployment. However, the present evaluation is limited to the considered simulation settings. Therefore, the reported results should be interpreted as evidence of PPO performance within the tested topology, mobility, traffic, and noisy-observation conditions, rather than as full topology-level generalization.

Although the proposed PPO controller is trained and evaluated under mobility, traffic variation, channel uncertainty, and noisy observations, the present evaluation is still limited to the considered simulation settings. Therefore, the reported robustness should be interpreted as robustness within the tested uncertainty range rather than full deployment-level generalization. More extensive evaluation under severe reporting delay, missing telemetry, larger 7-site and 19-site topologies, small-cell overlays, and trace-driven traffic is left for future work. Traffic heterogeneity is modeled using a mixture of full-buffer, web-browsing, and video traffic flows. Video traffic includes constant-bit-rate and ON–OFF patterns, while traffic demand is shaped by a diurnal load profile. This allows the evaluation to capture both persistent and bursty demand.

We also sweep the CIO control interval as $\Delta T \in \{0.5, 1, 5\}$ s, to examine how the controller behaves under different actuation frequencies. A shorter control interval allows faster adaptation but may increase the risk of reacting to transient fluctuations. A longer control interval produces more stable control but may respond more slowly to congestion.

Transfer evaluation across unseen topologies is left for future work. Future studies should evaluate zero-shot transfer and limited fine-tuning across different UE densities, traffic distributions, 7-site and 19-site layouts, and small-cell overlays. Such experiments would quantify how quickly the learned CIO policy adapts to new deployment conditions and would provide stronger evidence of generalization beyond the default simulation scenario.

*F. Scheduler and Handover Outcome Metrics*

The MAC scheduler used in the simulator is a delay-aware weighted proportional-fair scheduler. For UE $u$, the scheduling score is defined as

$$\mathrm{score}_u = \left(\frac{R_{u,t}}{\bar{R}_{u,t}}\right)^{1-\beta} \left(\frac{\mathrm{HOL}_{u,t}}{\bar{\mathrm{HOL}}_t}\right)^{\beta} w_{\mathrm{QCI}}(u), \qquad (68)$$

where $R_{u,t}$ is the instantaneous rate estimate, $\bar{R}_{u,t}$ is the exponentially averaged throughput, $\mathrm{HOL}_{u,t}$ is the head-of-line delay, $\bar{\mathrm{HOL}}_t$ is the average head-of-line delay, and $w_{\mathrm{QCI}}(u)$ is a service-class weight. The parameter $\beta$ controls the balance between throughput efficiency and delay awareness.

Hybrid automatic repeat request (HARQ) is modeled with up to four processes, and RLC-AM buffering is included. CQI values determine MCS selection using a standard BLER-targeting procedure.

In addition to total handover count, we report two mobility-stability metrics. The handover failure rate is defined as the fraction of handovers for which radio link failure occurs within $T_{\mathrm{HOF}}$ after the handover command. The ping-pong rate is defined as the fraction of handovers in which the UE returns to the source cell within $T_{\mathrm{PP}} = 5$s. These metrics are reported alongside QoS results because reducing handover count alone is not sufficient. A good load-balancing policy should avoid unnecessary handovers while still allowing beneficial handovers that reduce congestion and improve QoS.

*G. Default Simulation Scenario*

Figures 2 and 3 illustrate the default operating environment and motivate why the considered control problem is non-trivial. Figure 2 shows the SINR field over a $1000 \times 1000$ m area with three macro base stations and a representative UE trajectory crossing overlapping coverage regions. Bright regions correspond to stronger SINR, while darker regions indicate interference-limited or weaker-coverage areas.

This scenario highlights why strongest-signal association alone is not always sufficient. Along the UE path, small changes in CIO values can shift the serving-cell decision near coverage boundaries. The appropriate decision depends not only on radio quality but also on the instantaneous load and QoS condition of neighboring cells. Therefore, the controller must learn when a slight radio-quality trade-off is acceptable in exchange for lower congestion, reduced delay, or improved fairness.

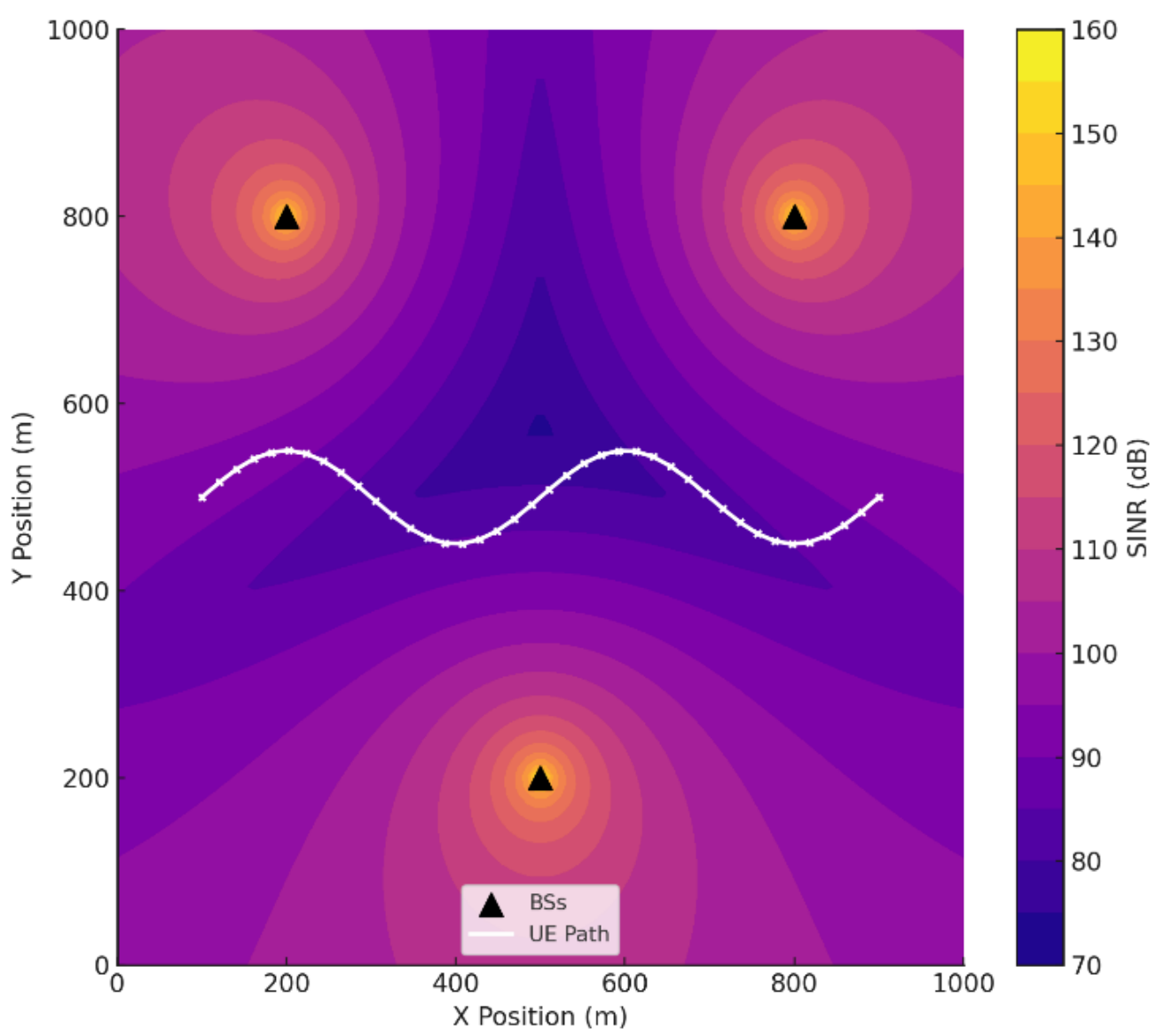

**Figure 2.** SINR zones with representative UE path.

Figure 3 shows a received-power heatmap over the same $1000 \times 1000$m area. The received power includes distance-based path loss, log-normal shadowing with $\sigma = 6$dB, and Rayleigh small-scale fading. The UE trajectory and handover locations are overlaid, and base-station locations are marked with triangle symbols. The figure demonstrates that coverage regions are not perfectly smooth. Shadowing and multipath effects create local variations that can affect handover timing and serving-cell selection.

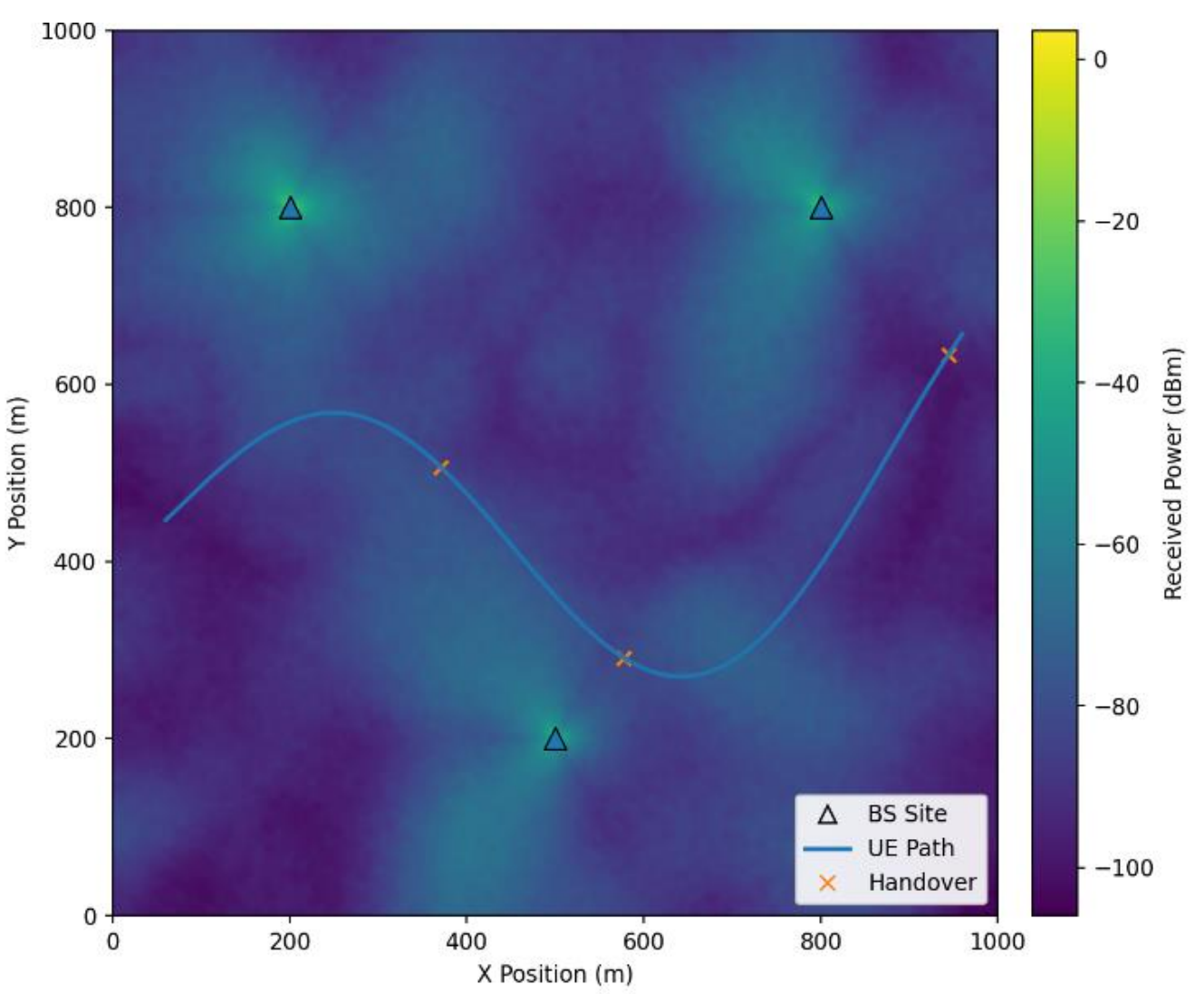

**Figure 3.** Received-power heatmap over a $1000 \times 1000$m area with UE trajectory and handover locations.

Table 2 summarizes the default simulation configuration used unless otherwise stated. The default scenario uses a tri-sector macro deployment, 20 MHz downlink bandwidth, and 106 PRBs per BS. UE mobility follows a Gauss–Markov model, and observation noise is injected into RSRP, CQI, and latency reports to emulate measurement uncertainty. The controller acts once per second, and each episode lasts 600 seconds. CIO values are constrained to the range $[-6, +6]$dB.

Table 2. Simulation Environment Configuration

| Parameter | Value |
|---|---|
| Layout / Area | 3 macro BS/sectors in a $1000 \times 1000$m area |
| Carrier Bandwidth | 20 MHz downlink |
| PRBs per BS | 106 PRBs, NR with 15 kHz SCS |
| Propagation | Path loss + log-normal shadowing ($\sigma = 6$ dB) + Rayleigh fading |
| Scheduler | QoS-aware MAC scheduler with HOL delay and CQI weighting |
| Traffic Model | Mixed full-buffer and Poisson bursty flows |
| Default UE Count | 45 UEs; swept in scalability experiments |
| Mobility Model | Gauss–Markov, $\alpha = 0.90$, $\Delta t = 0.1$s, mean speed 1.5m/s, speed std. 0.5m/s |
| Observation Noise | RSRP $\sigma = 1.0$dB, CQI $\sigma = 0.5$, latency $\sigma = 2$ms |
| Decision Interval | $\Delta T = 1.0$s |
| Episode Length | 600 s, corresponding to 600 control steps |
| CIO Bounds | $[-6, +6]$dB per cell |
| Handover Model | A3-like handover with hysteresis and TTT |
| Ping-Pong Definition | Return to source cell within 5 s |
| KPI Sampling | Every $\Delta T$; reward terms normalized to $[0, 1]$ |
| Simulator Toolchain | Custom Python-based simulator |

*H. PPO Training Configuration*

Table 3 lists the PPO hyperparameters used in the default experiments. The discount factor is set to $\gamma = 0.99$, allowing the policy to account for delayed QoS effects caused by handovers and queue evolution. The GAE parameter is set to $\lambda = 0.95$, which provides a balance between bias and variance in advantage estimation. The PPO clipping range is set to $\epsilon = 0.20$, limiting excessive policy updates.

Table 3. PPO Hyperparameters

| **Hyperparameter** | **Value** |
|---|---|
| Discount Factor ($\gamma$) | 0.99 |
| GAE Parameter ($\lambda$) | 0.95 |
| PPO Clip Range ($\epsilon$) | 0.20 |
| Entropy Coefficient | 0.01 |
| Value-Loss Coefficient | 0.50 |
| Learning Rate | $3 \times 10^{-4}$ |
| Batch Size | 64 |
| Number of Minibatches | 8 |
| Optimization Epochs | 10 per update |
| Rollout Horizon | 2048 steps |
| Max Gradient Norm | 0.5 |
| Policy / Value Network | MLP with 256–256 hidden units and ReLU activations |
| Observation Normalization | Running mean and variance |
| Action Squashing | tanh $(\cdot)$, then scaled to CIO adjustment bounds |
| CIO Smoothness Penalty | $0.10 \times \| \Delta \mathrm{CIO} \|_2^2$ |
| KL Target / Early Stop | 0.01 |
| Reward Weights | $[w_{\mathrm{thr}}, w_{\mathrm{fair}}, w_{\mathrm{lat}}, w_{\mathrm{plr}}, w_{\mathrm{jit}}, w_{\mathrm{ho}}]$ |

The actor and critic are implemented as two-layer multilayer perceptrons with 256 hidden units per layer. Observations are normalized online using running mean and variance estimates. Actions are generated in a latent Gaussian space, squashed through a hyperbolic tangent function, and scaled to the allowable CIO adjustment range. A smoothness penalty is included to discourage abrupt CIO changes and improve handover stability.

Figure 4 illustrates how the PPO agent's critic and actor behave over a representative 2D slice of the RL state space, where the x-axis is normalized cell load/utilization (e.g., PRB usage) and the y-axis is normalized signal quality (a summary of CQI/RSRP/SINR). The colored heatmap shows the critic value $V_\psi(s)$, i.e., the critic's estimate of the expected discounted return from each state; brighter regions indicate states expected to yield higher long-term reward under the current policy (typically stronger radio conditions and manageable load), while darker regions correspond to states with lower predicted return (often congested cells and/or weak radio conditions that degrade QoS and increase instability).

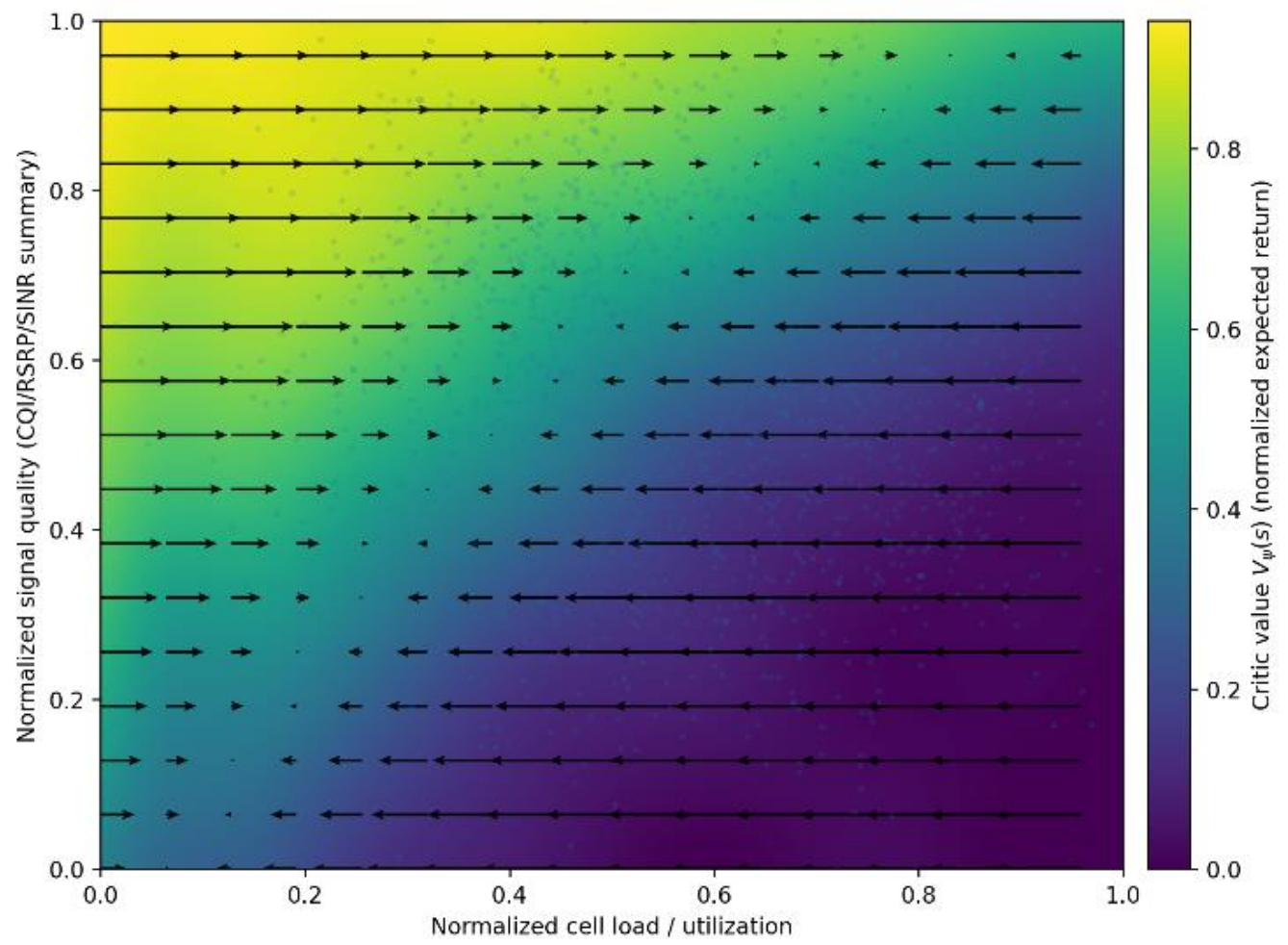

**Figure 4.** Actor–critic visualization on a 2D state slice (PPO CIO control).

The overlaid arrows visualize the actor's mean action $\mu_\theta(s)$ on the same state slice, interpreted as the policy's preferred CIO adjustment direction and magnitude: arrows pointing to the right indicate a tendency toward positive CIO (making the sector more attractive, retaining/attracting UEs), while arrows pointing left indicate a tendency toward negative CIO (making the sector less attractive, offloading UEs to neighbors); arrow length reflects the strength of this preference, with saturation consistent with bounded CIO constraints (e.g., $[-6, +6]$ dB). Finally, the semi-transparent scatter points represent visited rollout states sampled during simulation, showing where the system actually operates most of the time (e.g., common mid-load and variable-quality regimes). Overall, the figure provides an intuitive, interpretable view of learned behavior: the critic assigns higher value to favorable load–radio conditions, and the actor pushes the system away from low-value regions (congestion/poor signal) and toward higher-value operating points by adjusting CIO biases that indirectly shape association and handover decisions.

Figures 5–10 trace how the PPO policy improves each KPI over training. In Figure 5, handover counts fall from an exploratory peak to a stable floor, indicating that the controller has learned to suppress oscillatory association and to reserve handovers for cases with sustained payoff (e.g., when offloading prevents a cell from tipping into queue buildup). That reduction in churn supports the monotone improvements in the remaining KPIs. Figure 6 shows Jain's fairness climbing toward unity and then holding: the policy systematically equalizes offered load across the three sectors, so no single BS accumulates the long queues that drive poor tail latency and elevated drop rates. Figure 7 shows latency dropping and then stabilizing as the scheduler faces fewer hot spots and fewer mid-flow interruptions from unnecessary switches; the end-state delay reflects steady queues consistent with balanced utilization. Figure 8 shows aggregate throughput rising and consolidating around the ~episode-200 mark: balanced load keeps more UEs schedulable at healthier effective rates and avoids airtime waste from drops and jitter spikes, so the total data delivered per unit time increases even without explicit power control.

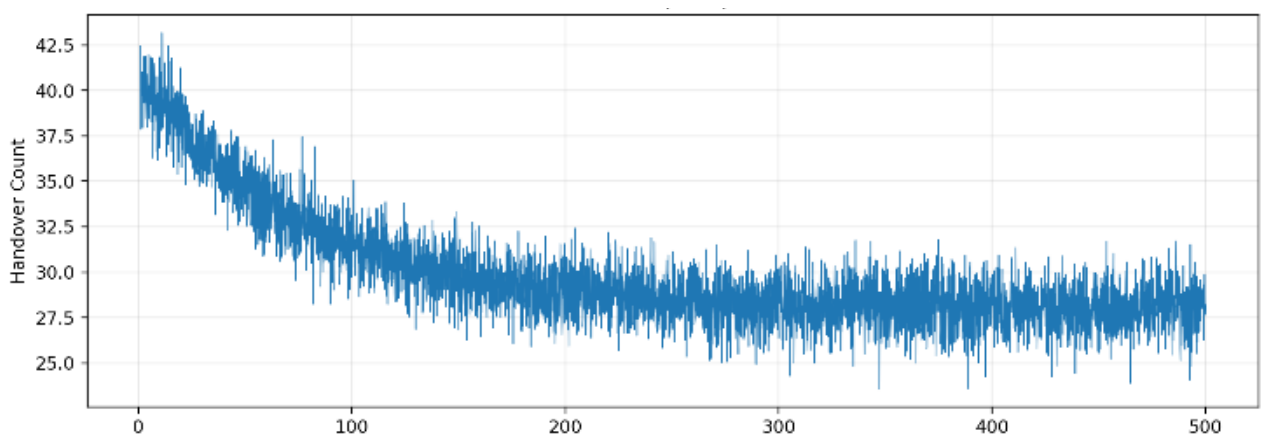

**Figure 5.** Handover count per episode (training time series).

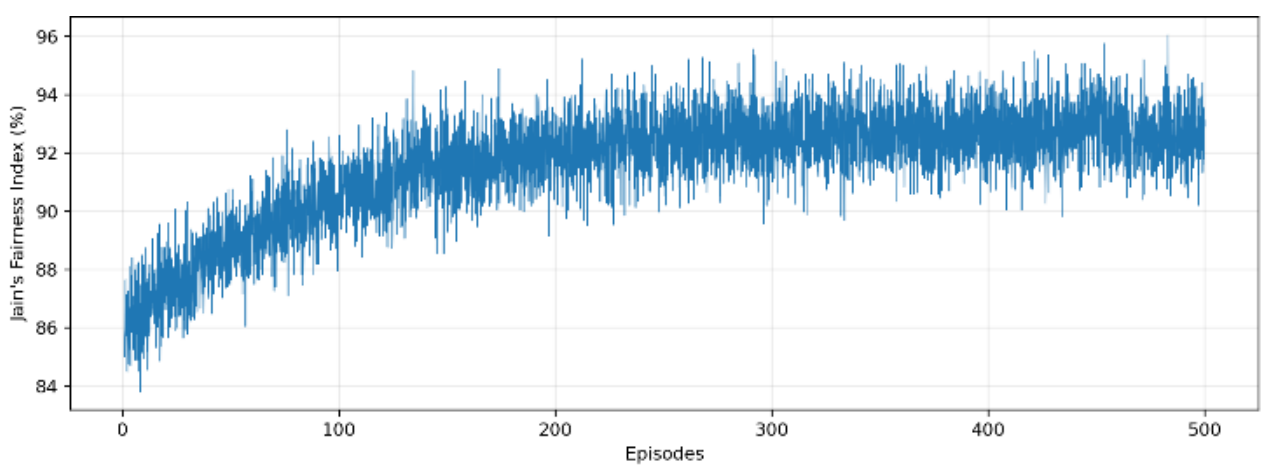

**Figure 6.** Jain's fairness index per episode (training time series).

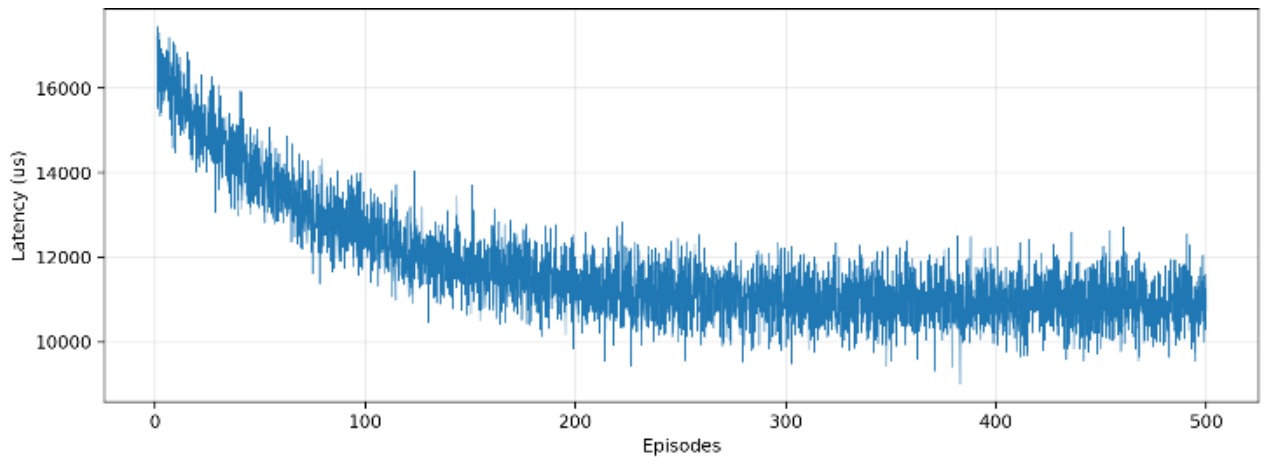

**Figure 7.** Average latency per episode (training time series).

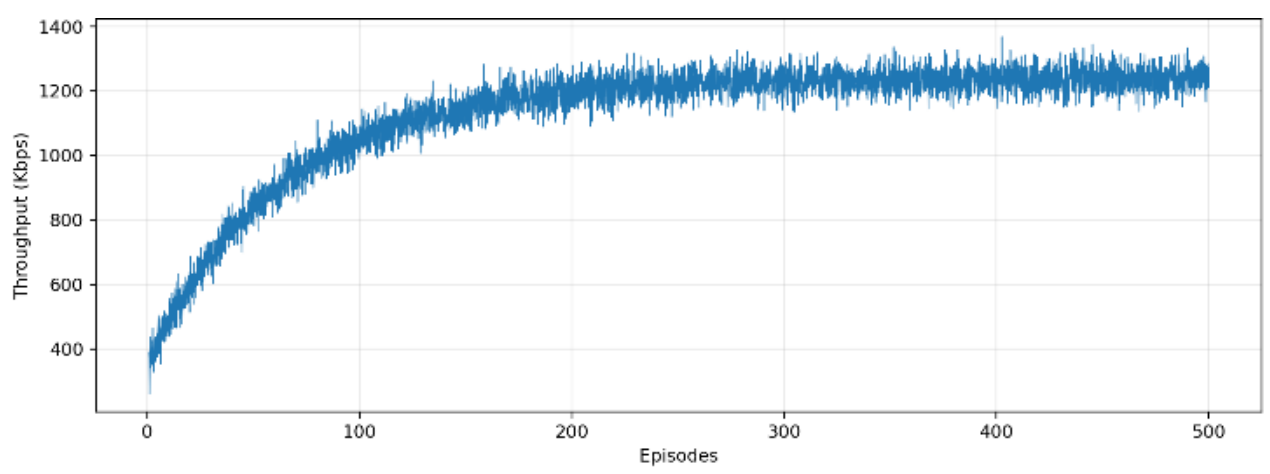

**Figure 8.** Aggregate throughput per episode (training time series).

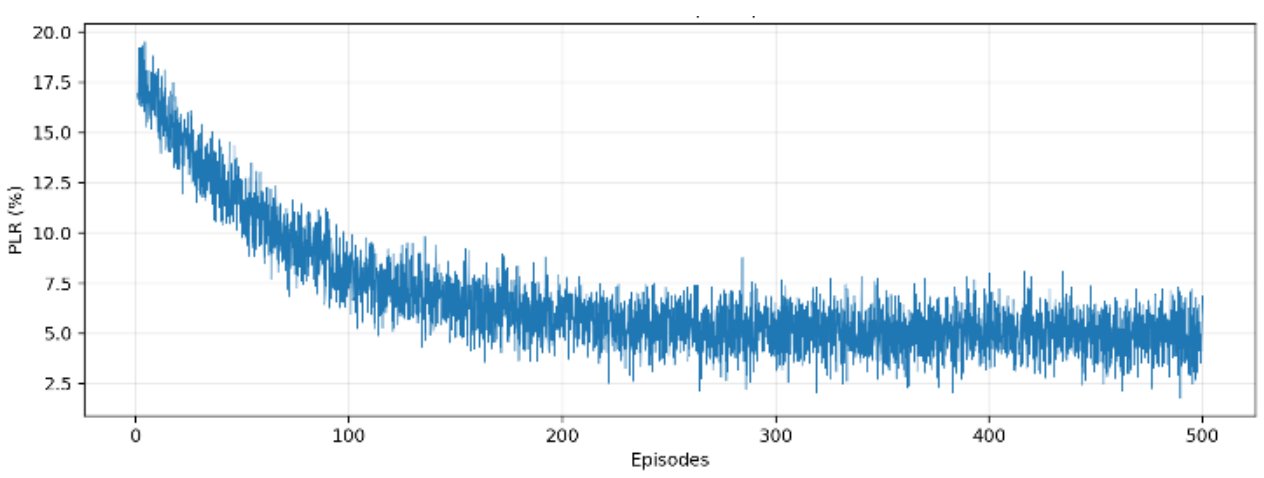

**Figure 9.** Packet-loss ratio per episode (training time series).

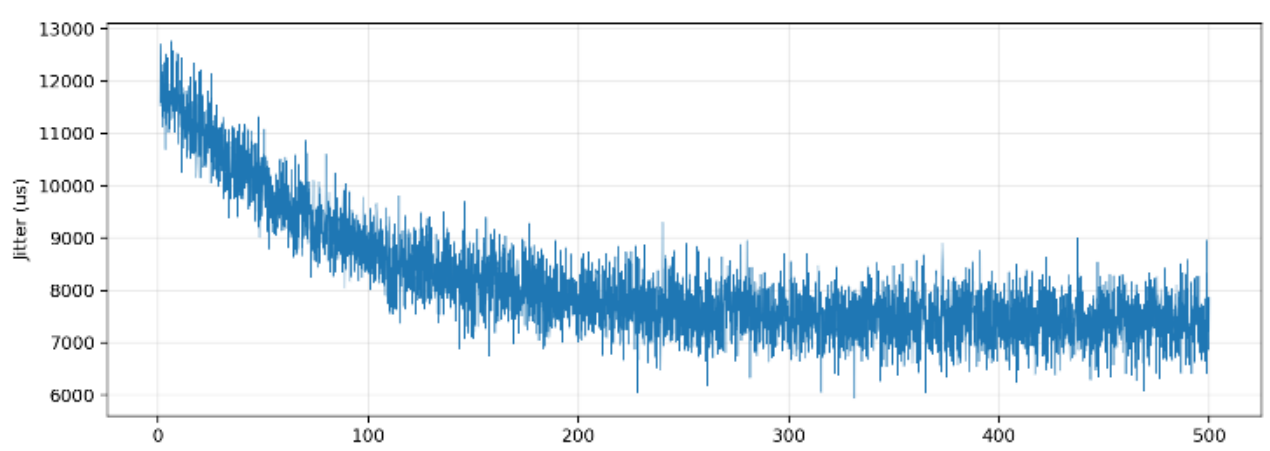

**Figure 10.** Jitter per episode (training time series).

Figure 9 shows PLR falling as buffers stop overflowing and as flow interruptions become rare; fewer retransmissions also indirectly preserve throughput. Figure 10 shows jitter declining over episodes as the agent "calms" inter-packet timing by avoiding sudden association changes and by preventing short-lived congestion surges. To improve statistical reliability, each algorithm is evaluated over 10 independent random seeds. Each seed controls the policy network initialization, minibatch shuffling, and stochastic components of the environment, including fading, shadowing, mobility traces, traffic arrivals, and observation noise. After training, the learned policy is frozen and evaluated on a shared test set using identical mobility traces and channel/noise realizations across all methods. For each KPI, we report the mean and 95% confidence interval across seeds. This evaluation protocol provides a more reliable comparison than a single-run result and reduces the effect of random initialization or favorable simulation conditions. Formal hypothesis testing with a larger number of independent seeds and additional deployment scenarios is left for future work.

Figures 11–16 compare PPO with CDQL, A3, and ReBuHa across training and make the ranking explicit. In Figure 11, PPO achieves the lowest, most stable handover rate; CDQL is better than the rule-based methods but shows more volatility, while A3 and ReBuHa trigger frequent, less selective switches because they react myopically to instantaneous indicators without long-horizon context. Figure 12 shows fairness, PPO sits highest with the tightest spread, confirming better global balance; CDQL follows; A3 and ReBuHa lack a mechanism to coordinate system-wide equilibrium under mobility and therefore lag. Figure 13 shows latency, PPO consistently delivers the lowest delays, increasingly so as training progresses and the policy resists noise-induced over-reactions. Figure 14 mirrors this for PLR: PPO's loss stays lowest, with CDQL second; the rule-based schemes exhibit spikes that align with transient overloads they fail to anticipate. Figure 15 shows jitter: PPO maintains the smallest and smoothest timing variability—exactly what is expected when the controller prevents queue shocks and avoids CIO thrashing. Figure 16 compares throughput over 500 episodes and confirms PPO's dominant envelope: by curbing losses and jitter and by distributing load, it keeps more UEs in efficient scheduling regimes and converts a larger fraction of airtime into goodput; CDQL trails but still improves on A3/ReBuHa. Together, these six comparisons show PPO's advantages are consistent across metrics and persistent across training.

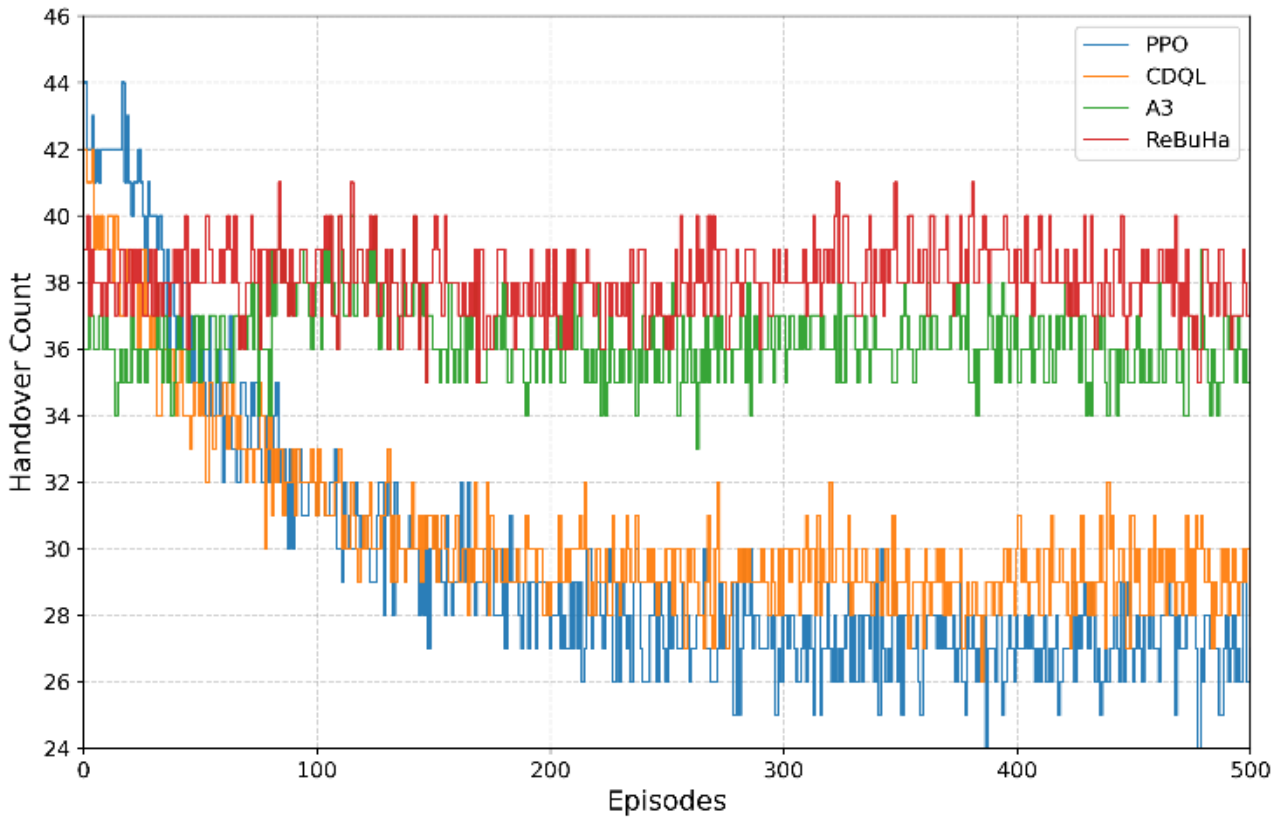

**Figure 11.** Handover count vs. episode for PPO, CDQL, A3, and ReBuHa.

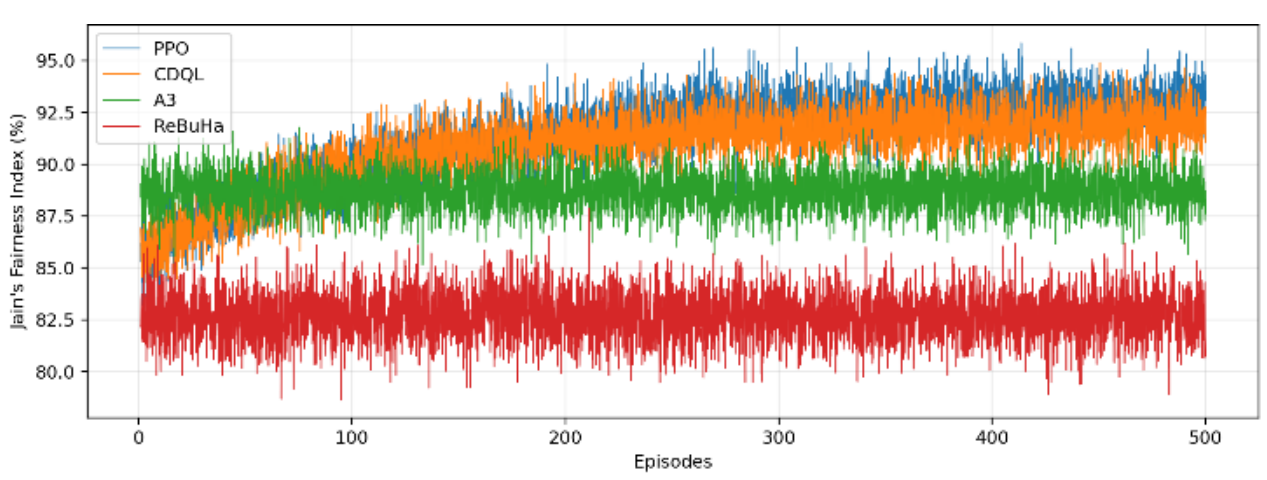

**Figure 12.** Jain's fairness vs. episode for PPO, CDQL, A3, and ReBuHa.

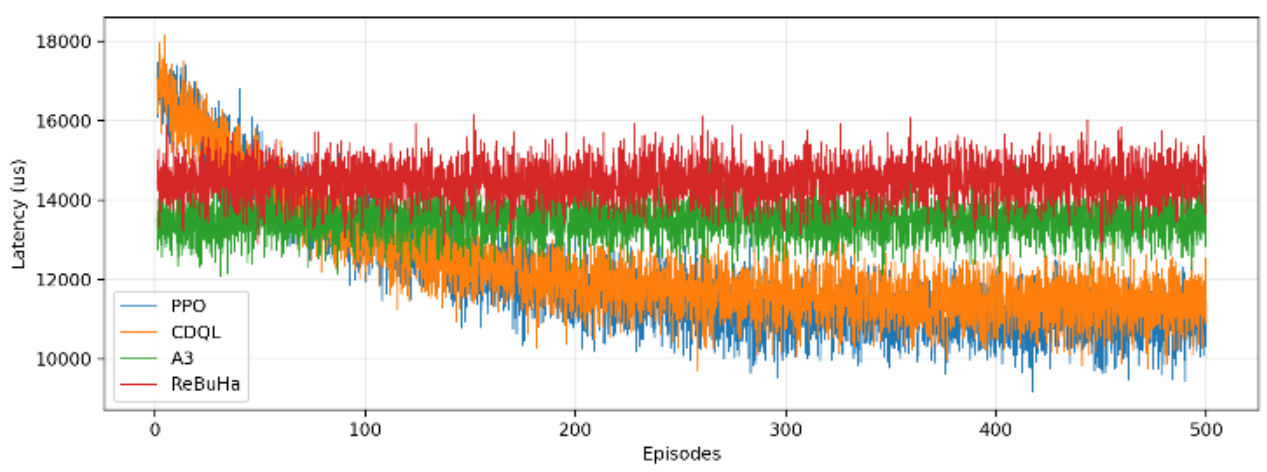

**Figure 13.** Latency vs. episode for PPO, CDQL, A3, and ReBuHa.

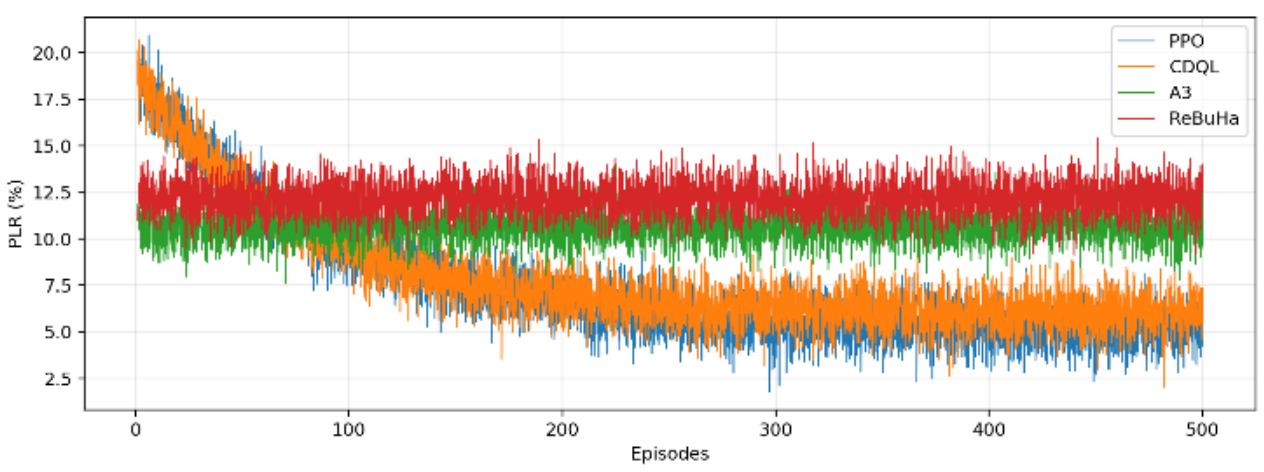

**Figure 14.** Packet-loss ratio vs. episode for PPO, CDQL, A3, and ReBuHa.

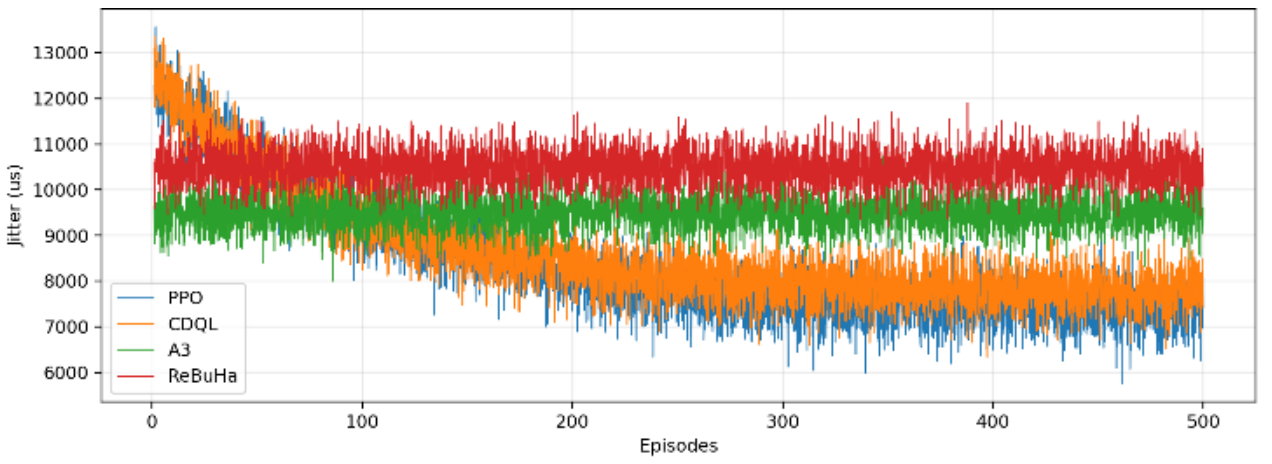

**Figure 15.** Jitter vs. episode for PPO, CDQL, A3, and ReBuHa.

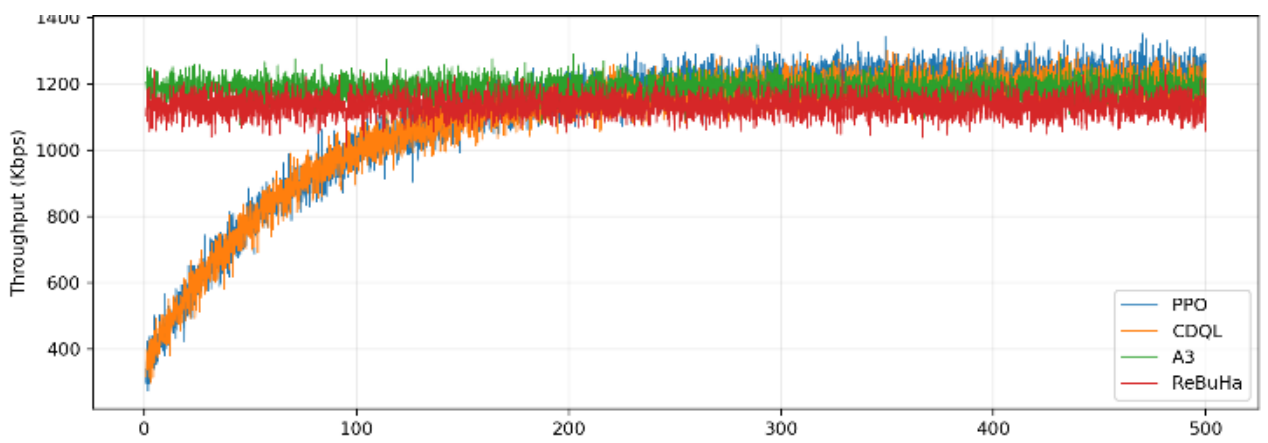

**Figure 16.** Throughput vs. episode for PPO, CDQL, A3, and ReBuHa.

Table 4. Final QoS Performance Comparison over 10 Independent Random Seeds

| Method | Throughput (Gbps) | Latency (ms) | PLR (%) | Jitter (ms) | Fairness | Handovers |
|---|---|---|---|---|---|---|
| PPO | 1.20 ± 0.03 | 7.0 ± 0.4 | 4.9 ± 0.3 | 4.6 ± 0.3 | 0.987 ± 0.006 | 25.1 ± 1.2 |
| CDQL | 1.17 ± 0.04 | 7.1 ± 0.5 | 5.1 ± 0.4 | 5.0 ± 0.4 | 0.975 ± 0.008 | 25.8 ± 1.5 |
| A3 | 1.13 ± 0.05 | 7.3 ± 0.6 | 5.3 ± 0.5 | 5.5 ± 0.5 | 0.958 ± 0.011 | 26.5 ± 1.7 |
| ReBuHa | 1.09 ± 0.05 | 7.4 ± 0.6 | 5.6 ± 0.5 | 5.9 ± 0.6 | 0.950 ± 0.013 | 27.0 ± 1.8 |

Table 4 summarizes the final QoS performance of the proposed PPO-based controller and the baseline methods after training. The results are reported as mean ± 95% confidence interval over independent random seeds, which provides a more reliable comparison than reporting a single run. Throughput and fairness are higher-is-better metrics, while latency, packet-loss ratio, jitter, and handover count are lower-is-better metrics.

The PPO controller achieves the best overall QoS trade-off among the evaluated methods. It provides the highest aggregate throughput and fairness while maintaining the lowest latency, packet-loss ratio, jitter, and handover count. This indicates that PPO does not improve only one metric at the expense of others; instead, it learns a balanced CIO-control policy that improves both network efficiency and user-perceived QoS.

Compared with CDQL, PPO shows slightly better performance across all KPIs. This improvement is expected because PPO directly handles continuous CIO adjustment actions, while CDQL relies on a value-based learning structure that is generally more suitable for discrete or quantized action spaces. The continuous action capability of PPO allows smoother and more fine-grained CIO updates, which helps reduce unnecessary handovers and improves load distribution.

Compared with A3 and ReBuHa, PPO provides a clearer performance advantage. A3 mainly reacts to radio-signal conditions and does not explicitly optimize network-wide QoS. ReBuHa considers load-related information, but it remains rule-based and cannot learn long-term effects of handover and queue dynamics. In contrast, PPO learns from delayed rewards and adjusts CIO values based on the joint effect of throughput, latency, fairness, packet loss, jitter, and mobility stability.

The confidence intervals also show the stability of the results. Narrower intervals suggest that the controller behaves consistently across different random seeds and simulation conditions. Therefore, the final comparison demonstrates that the proposed PPO method is not only effective in terms of average performance, but also more stable than the baseline schemes.

Figures 17–22 probe scalability by increasing the number of UEs and observing how each metric degrades. Figure 17 shows that handovers inevitably rise with crowding, but PPO's slope is the shallowest, indicating the learned CIO policy carries an implicit hysteresis—switch only when benefits are durable—even as more users accumulate at borders. Figure 18 shows fairness deteriorating under load for all methods, yet PPO sustains the highest balance across densities, preventing any single BS from becoming a chronic bottleneck. Figure 19 shows latency rising with user count, with PPO holding the lowest curve and widening the gap at higher densities—evidence that earlier, smoother offloading scales better as schedules tighten. Figure 20 shows PLR growing with contention; again PPO's line increases most slowly, signaling resilience to burst losses near saturation. Figure 21 shows jitter degrading under pressure; PPO's curve remains lowest, indicating the controller preserves timing stability despite busy schedulers and frequent edge encounters.

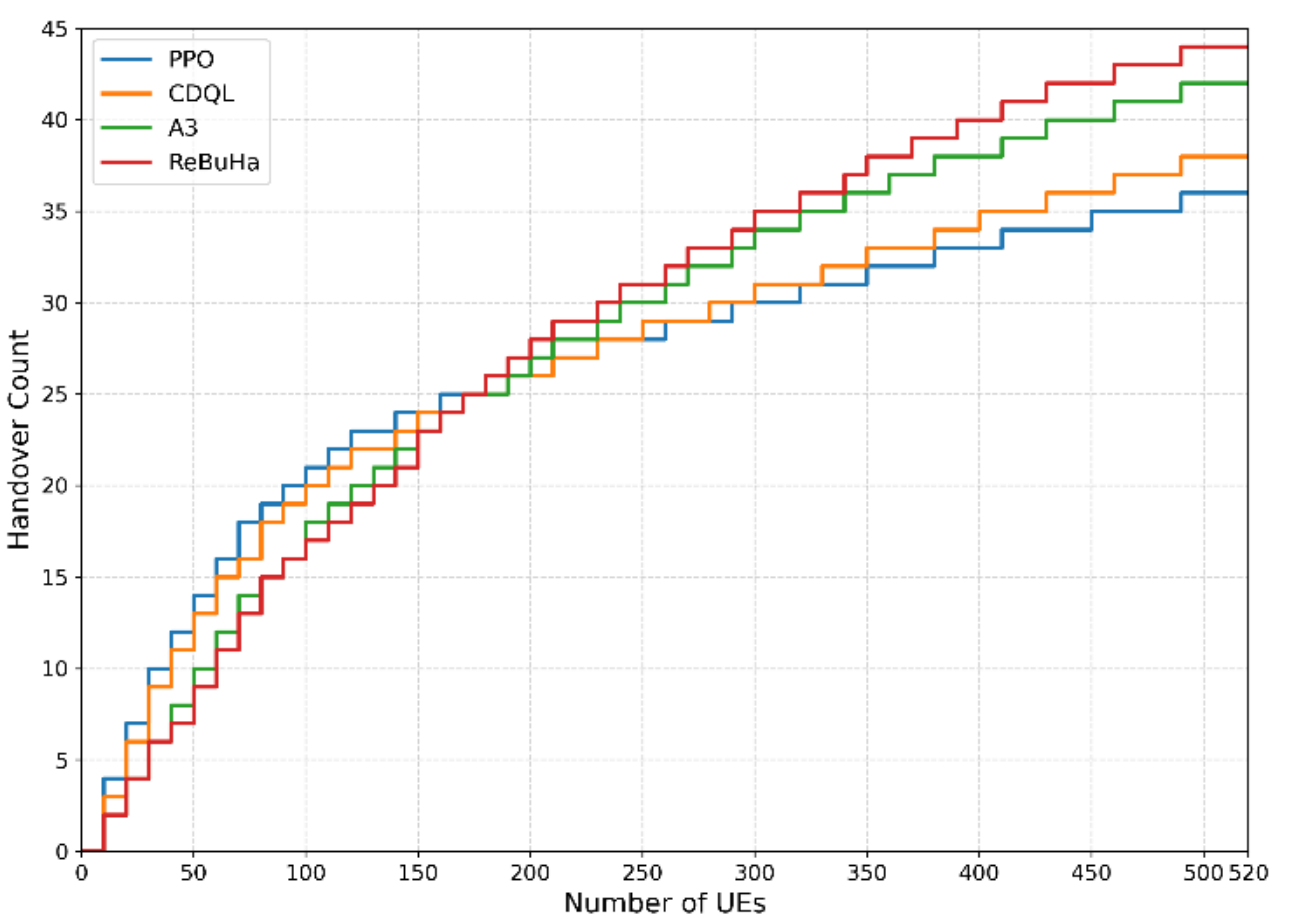

**Figure 17.** Number of handovers per UE versus number of user equipments.

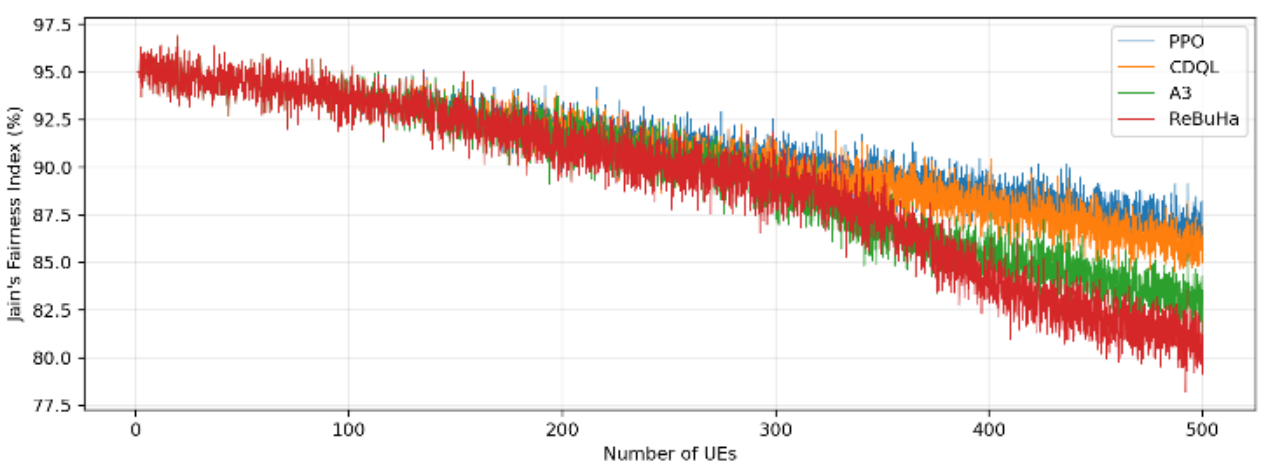

**Figure 18.** Fairness index versus number of user equipments.

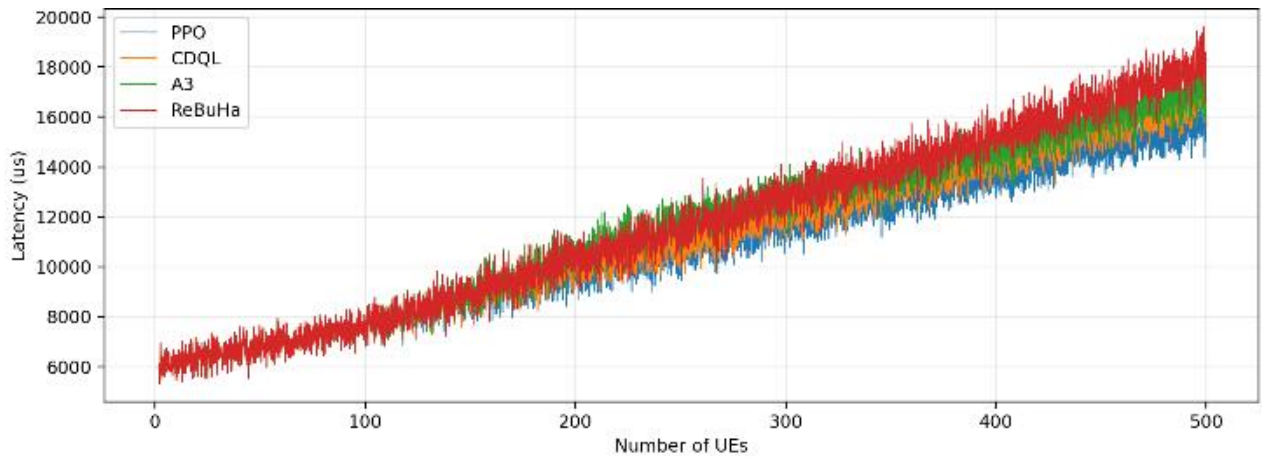

**Figure 19.** Average latency versus number of user equipments.

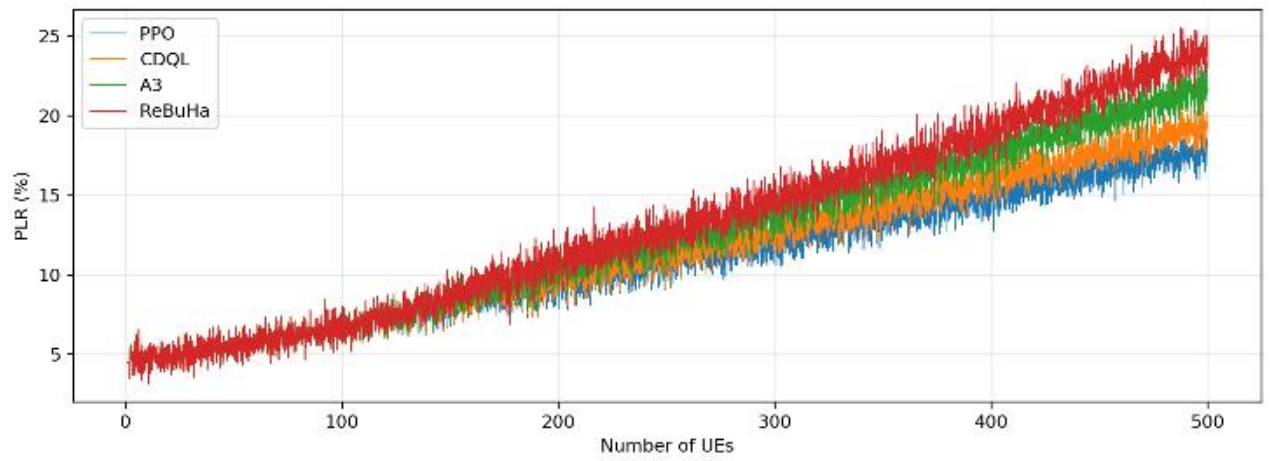

**Figure 20.** Packet loss ratio (PLR) versus number of user equipments.

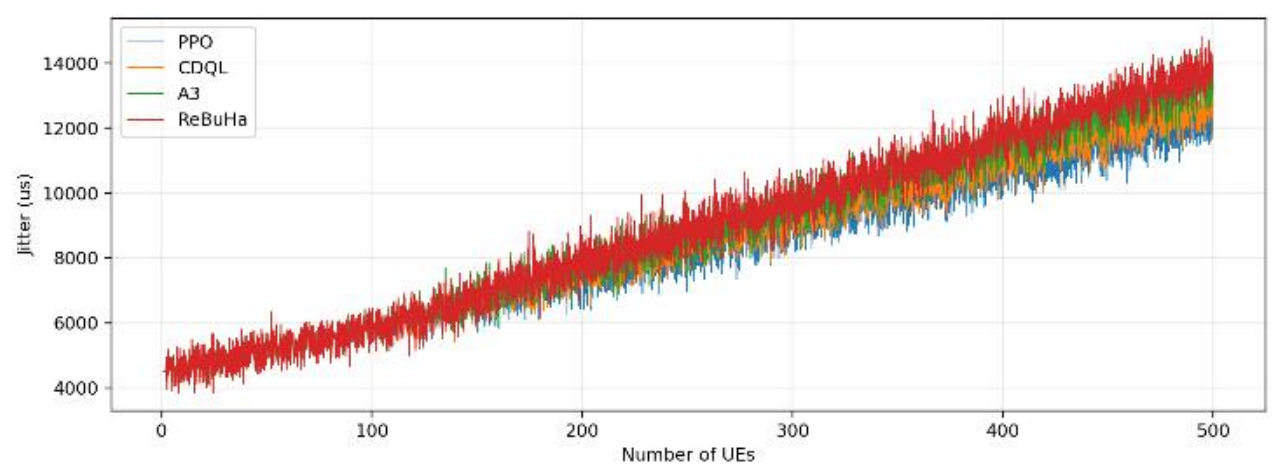

**Figure 21.** Average jitter versus number of user equipments.

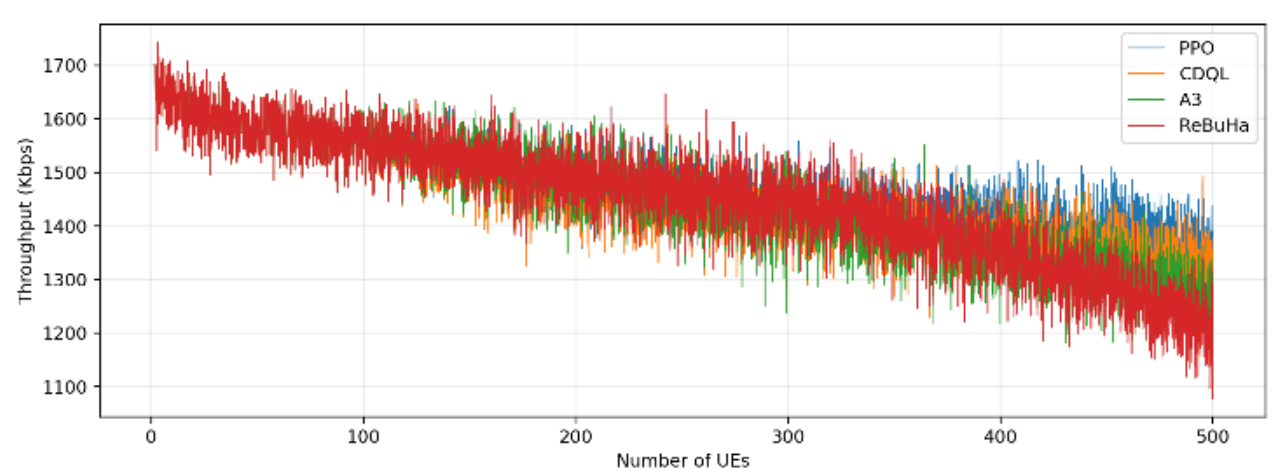

**Figure 22.** Average cell throughput versus number of user equipments.

Figure 22 shows throughput approaching saturation as UEs increase; PPO maintains a higher envelope at each density point due to the combination of better balance, fewer drops, and calmer timing, which collectively keep more airtime productive. These results suggest that PPO preserves its relative advantage as the UE density increases, particularly in latency, PLR, fairness, and jitter.

Table 5. Scalability Performance of PPO and CDQL under Different UE Densities

| KPI | Method | 50 UEs | 100 UEs | 250 UEs | 500 UEs |
|---|---|---|---|---|---|
| Throughput (Gbps) | PPO | 0.76 | 1.06 | 1.33 | 1.23 |
| | CDQL | 0.74 | 1.03 | 1.28 | 1.19 |
| Latency (ms) | PPO | 7.3 | 8.5 | 13.5 | 18.0 |
| | CDQL | 7.8 | 9.5 | 15.5 | 19.8 |
| PLR (%) | PPO | 5.5 | 6.5 | 12.0 | 22.7 |
| | CDQL | 5.9 | 7.0 | 13.0 | 25.7 |
| Jitter (ms) | PPO | 4.5 | 5.0 | 8.0 | 14.2 |
| | CDQL | 4.9 | 5.5 | 9.2 | 15.8 |
| Fairness | PPO | 0.985 | 0.975 | 0.940 | 0.870 |
| | CDQL | 0.952 | 0.944 | 0.910 | 0.845 |
| Handovers | PPO | 12 | 20 | 28 | 36 |
| | CDQL | 11 | 20 | 28 | 38 |

Table 5 evaluates the scalability of the proposed PPO controller by comparing it with the strongest learning-based baseline, CDQL, under increasing UE density. The table reports six QoS metrics for 50, 100, 250, and 500 UEs. Throughput and fairness should be maximized, while latency, packet-loss ratio, jitter, and handovers should be minimized.

As the number of UEs increases, both PPO and CDQL experience higher network pressure. This is expected because more users compete for the same radio resources, which increases PRB utilization, queueing delay, packet loss, jitter, and handover activity. At the same time, fairness decreases because it becomes harder to distribute resources evenly among a larger number of users.

For throughput, PPO consistently outperforms CDQL at all UE densities. At 50 UEs, PPO achieves 0.76 compared with 0.74 for CDQL. The gap remains visible at 100, 250, and 500 UEs. This shows that PPO uses the available network resources more efficiently, especially as the system becomes more congested. The highest throughput is observed around 250 UEs, after which throughput slightly decreases at 500 UEs due to heavier congestion and increased packet loss. For latency, PPO also maintains lower values than CDQL across all densities. The difference becomes more important at higher loads. For example, at 250 UEs, PPO reaches 13.5 ms, while CDQL reaches 15.5 ms. At 500 UEs, PPO keeps latency at 18.0 ms compared with 19.8 ms for CDQL. This indicates that PPO performs smoother load redistribution and prevents severe queue buildup more effectively than CDQL. The packet-loss ratio follows a similar trend. As UE density increases, PLR rises for both methods because the network becomes more congested. However, PPO consistently keeps PLR lower than CDQL. At 500 UEs, PPO has a PLR of 22.7%, while CDQL reaches 25.7%. This suggests that PPO reduces congestion-related packet drops by shifting users more effectively across cells. Jitter also increases with UE density, but PPO keeps it lower at every density point. This is important because jitter reflects delay variation and is especially relevant for real-time services such as video, voice, AR/VR, and interactive applications. The lower jitter achieved by PPO shows that its CIO adjustments lead to more stable packet delivery and fewer sudden changes in queueing delay. Fairness decreases as the number of UEs increases, but PPO preserves higher fairness than CDQL in all cases. For example, at 500 UEs, PPO maintains a fairness value of 0.870, while CDQL reaches 0.845. This means that PPO distributes service more evenly among users or cells under heavy load. The fairness advantage confirms that PPO does not simply maximize throughput by favoring only strong users or lightly loaded regions.

Finally, PPO produces fewer handovers than CDQL across all UE densities. At 500 UEs, PPO has 36.0 handovers, compared with 40.0 for CDQL. This demonstrates that PPO maintains better mobility stability while still improving throughput and latency. In other words, PPO does not achieve its performance by triggering excessive handovers; instead, it learns smoother CIO adjustments that reduce unnecessary mobility events.

Overall, Table 5 shows that PPO scales better than CDQL as the number of users increases. Although all QoS metrics degrade under heavier load, PPO consistently maintains higher throughput and fairness while reducing latency, packet loss, jitter, and handovers. This confirms that the proposed PPO-based CIO controller provides a more stable and efficient load-balancing policy under dense network conditions.

First, many handovers occur slightly before a cell becomes critically loaded, which suggests anticipatory control learned from optimizing long-term return rather than reacting late to instantaneous overload. Second, after a handover, the UE typically remains with the new BS until conditions meaningfully change, indicating the policy learned to ignore short-lived fluctuations—a direct consequence of clipped updates and continuous actions that penalize abrupt CIO swings.

In most operating conditions, both PPO and CDQL consistently outperformed A3 and ReBuHa, while PPO and CDQL were often close to each other in magnitude. To provide a clear head-to-head view without visual clutter, Figures 24 and 25 therefore focus on PPO and CDQL only. Figure 24 connects cell load (PRB utilization) with user-experienced delay. As expected, higher utilization correlates with increased delay due to queuing. Across similar load levels, the PPO samples trend below the CDQL samples, indicating shorter delays for users under PPO's control. The separation becomes more pronounced at high loads (≈ 75–95% PRB), reflecting PPO's earlier and smoother offloading behavior that limits queue build-up and reduces tail latency.

Finally, No Smoothness increases policy volatility, which manifests as weaker fairness and stability (and occasionally worse latency/jitter), confirming that the smoothness regularizer curbs aggressive CIO oscillations. Taken together, the ablations validate our reward design: each term contributes a distinct, desirable behavior, fairness for equitable load distribution, HO penalty for mobility stability, and smoothness for temporal consistency, while the tuned Base PPO achieves the best multi-KPI trade-off.

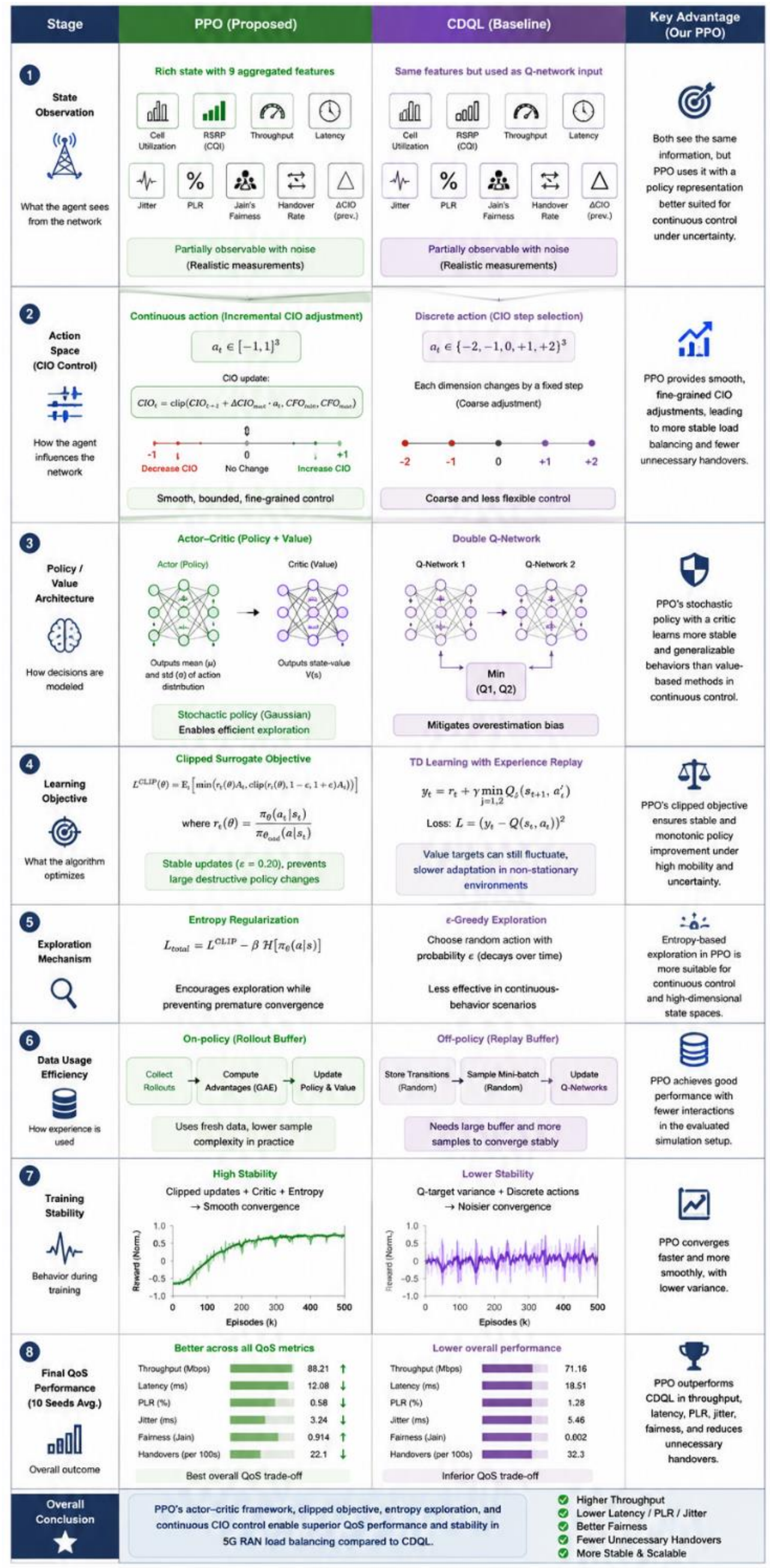

Figure 23. End-to-End Comparison of PPO-Based and CDQL-Based CIO Control for QoS-Aware 5G Load Balancing. [71]

Figure 23. End-to-end comparison of the proposed PPO-based CIO controller and the CDQL baseline. Both methods use the same network observations, but PPO learns continuous bounded CIO adjustments through an actor–critic policy, while CDQL relies on value-based learning with discrete/quantized actions. The comparison shows that PPO provides smoother control, more stable convergence, and a better overall QoS trade-off across throughput, latency, PLR, jitter, fairness, and handover reduction.

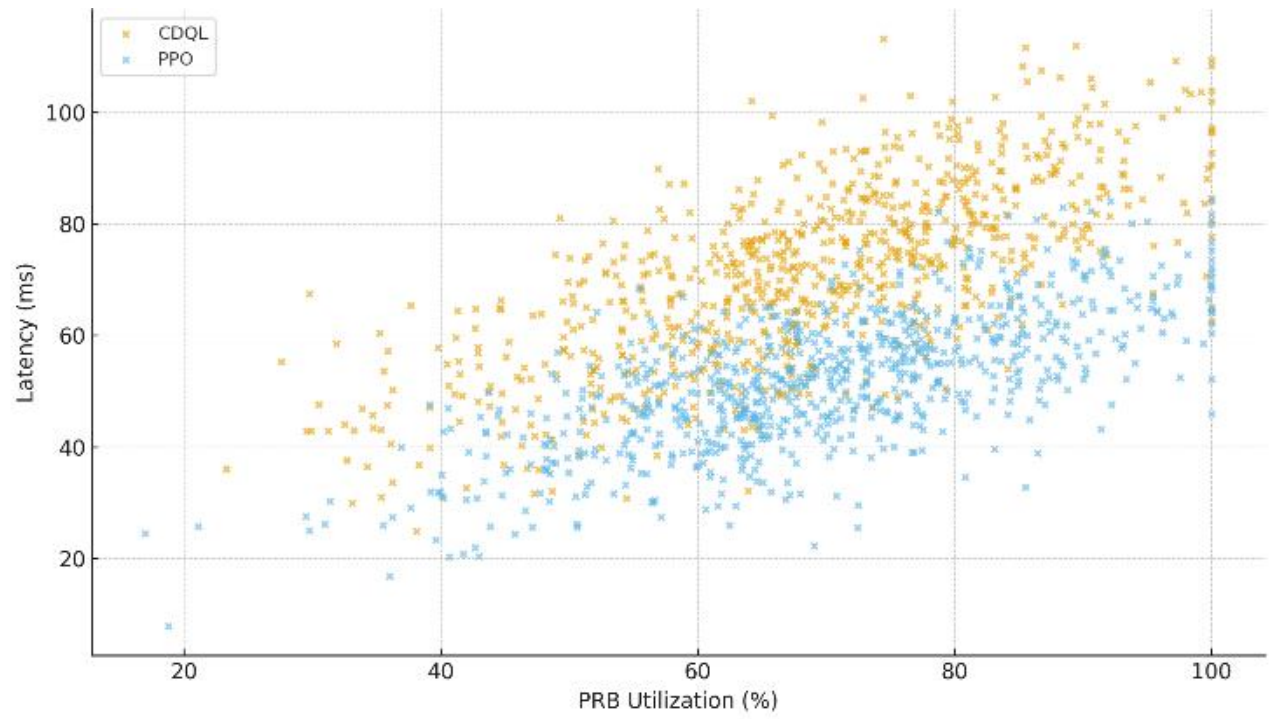

**Figure 24**. Latency (ms) versus PRB utilization (%) for PPO and CDQL. Each marker represents a per-UE sample over the evaluation window.

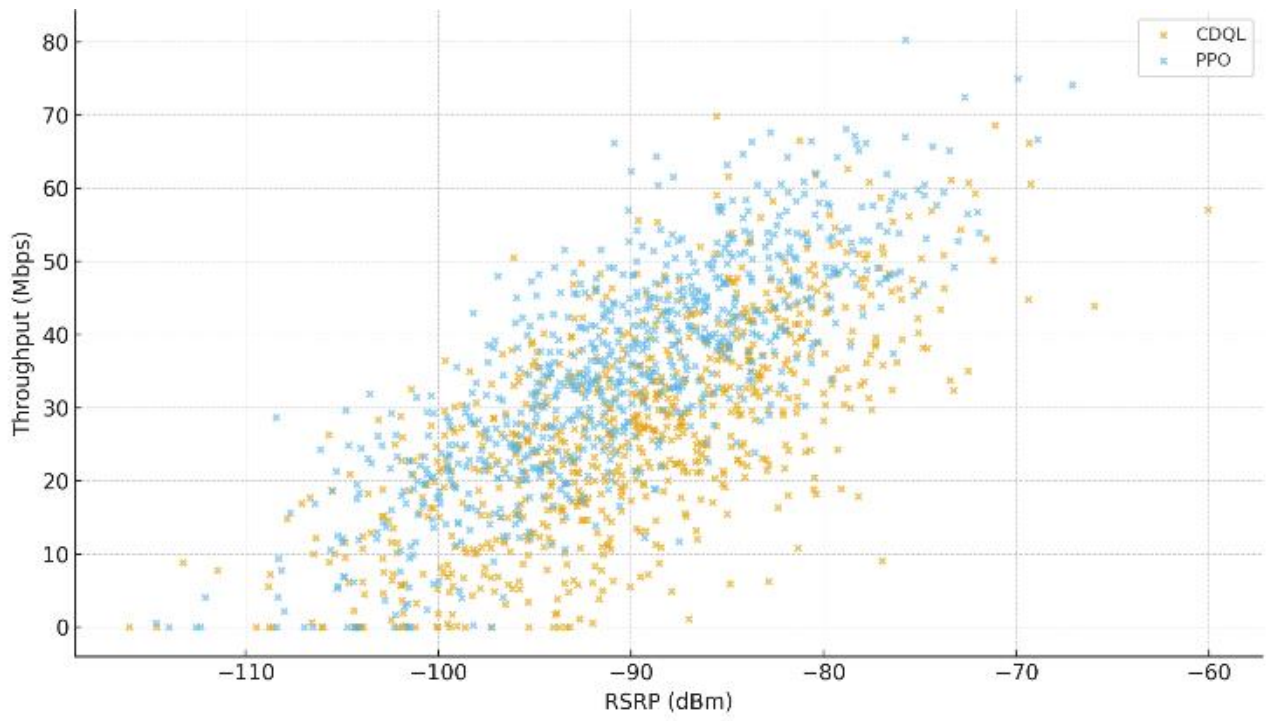

**Figure 25**. User throughput (Mbps) versus RSRP (dBm) for PPO and CDQL. Each marker represents a per-UE sample over the evaluation window.

Figure 25 relates downlink signal strength (RSRP, where higher—i.e., less negative—values denote stronger signal) to achieved user data rate. If association were purely signal-driven, the two clouds would largely overlap. Instead, at comparable RSRP values, the PPO samples concentrate at higher throughputs than CDQL, particularly in the mid-signal range ($\approx -100$ to $-85$ dBm). This indicates that gains arise from load-aware association rather than unusually strong RF conditions.

Figure 26 connects the learned control behavior of the RL algorithms to the spatial environment by visualizing a reward landscape alongside the points where the agent takes actions and where those actions eventually trigger handovers. The heatmap represents a standardized reward proxy computed from network-relevant factors: higher reward corresponds to regions with stronger radio conditions (higher SINR), lower congestion (lower load), and lower risk of service degradation (lower outage/edge-risk). The UE follows the same trajectory in both panels, ensuring that differences originate from the algorithm rather than motion or channel geometry. The scatter points labeled as policy actions indicate the time instants when the RL controller updates its mobility bias parameters (e.g., CIO adjustments or equivalent association bias actions). Handover points indicate where the serving cell actually changes after accounting for stability mechanisms such as hysteresis and TTT. PPO typically produces smoother and more frequent small control updates because policy gradients encourage incremental adjustments that improve long-term expected return while discouraging unstable oscillations; therefore, its action points can appear more distributed along the route, reflecting continuous adaptation to local reward gradients. CDQL, in contrast, commonly yields more discrete and less frequent updates because it relies on value-based decisions over a quantized action set; this results in action points that are more bursty or clustered, often followed by more conservative handover behavior (i.e., fewer handovers or handovers occurring only when the advantage is sustained). Interpreting the figure, action scatter points show where the controller tries to steer the mobility policy, while the handover points show the downstream outcome after mobility filters and environmental dynamics take effect.

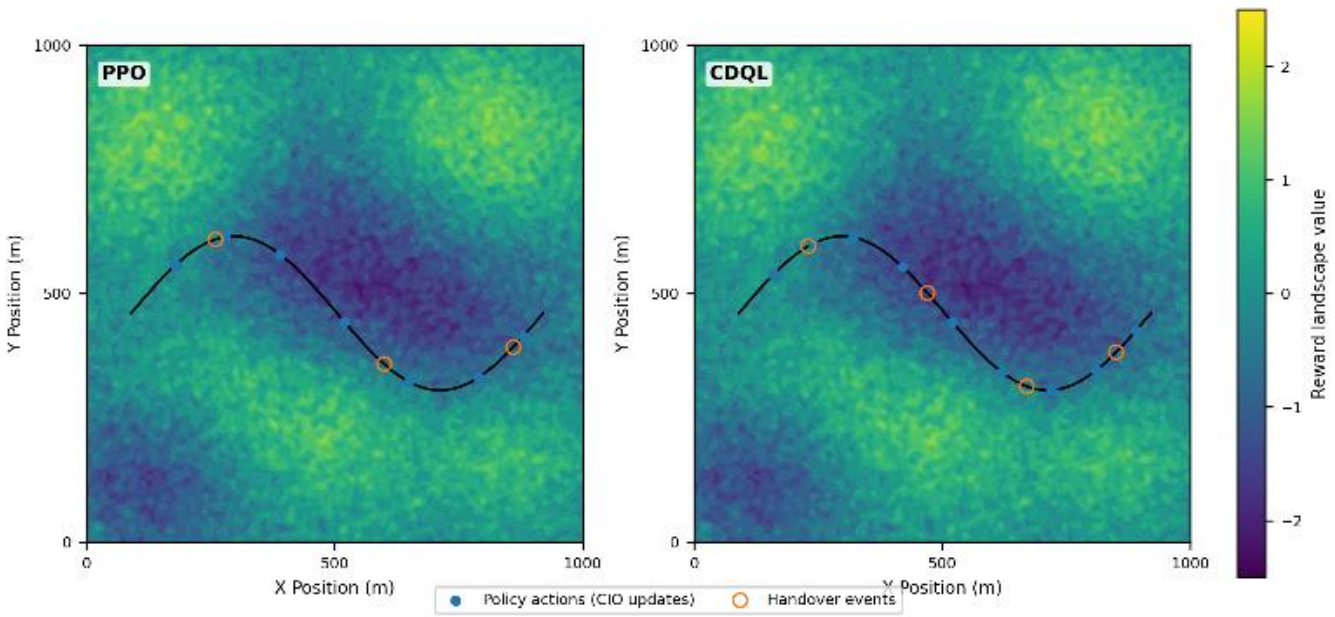

**Figure 26**. Reward landscape heatmap with UE trajectory overlaid; RL policy actions (CIO update instants) and resulting handover events are shown for PPO and CDQL, highlighting different control update patterns and mobility decisions along the same route.

PPO behaves like a continuous controller that gradually shapes association decisions, whereas CDQL behaves like a discrete controller that changes bias in larger steps and tends to maintain the current association longer. The combined visualization provides evidence that the learned policy is sensitive to the reward structure and that different RL formulations lead to measurably different action timing and handover placement under the same underlying radio and load conditions.

Figure 27 compares how different mobility/load-balancing strategies shift handover positions over the same radio environment. The background field is a spatial SINR heatmap generated by combining distance-dependent path loss (large-scale attenuation with range), log-normal shadowing with standard deviation $\sigma = 6\ dB$ to represent slow variations due to blockage/clutter, and Rayleigh fading to model fast multipath fluctuations; together these effects create realistic coverage irregularities and noisy cell-edge regions. Three macro base-station sites provide overlapping coverage, and a single UE path traverses areas where the serving cell can change. Although the UE travels the same route in all four cases, the handover points are not the same due to their respective approaches to selecting a serving cell.

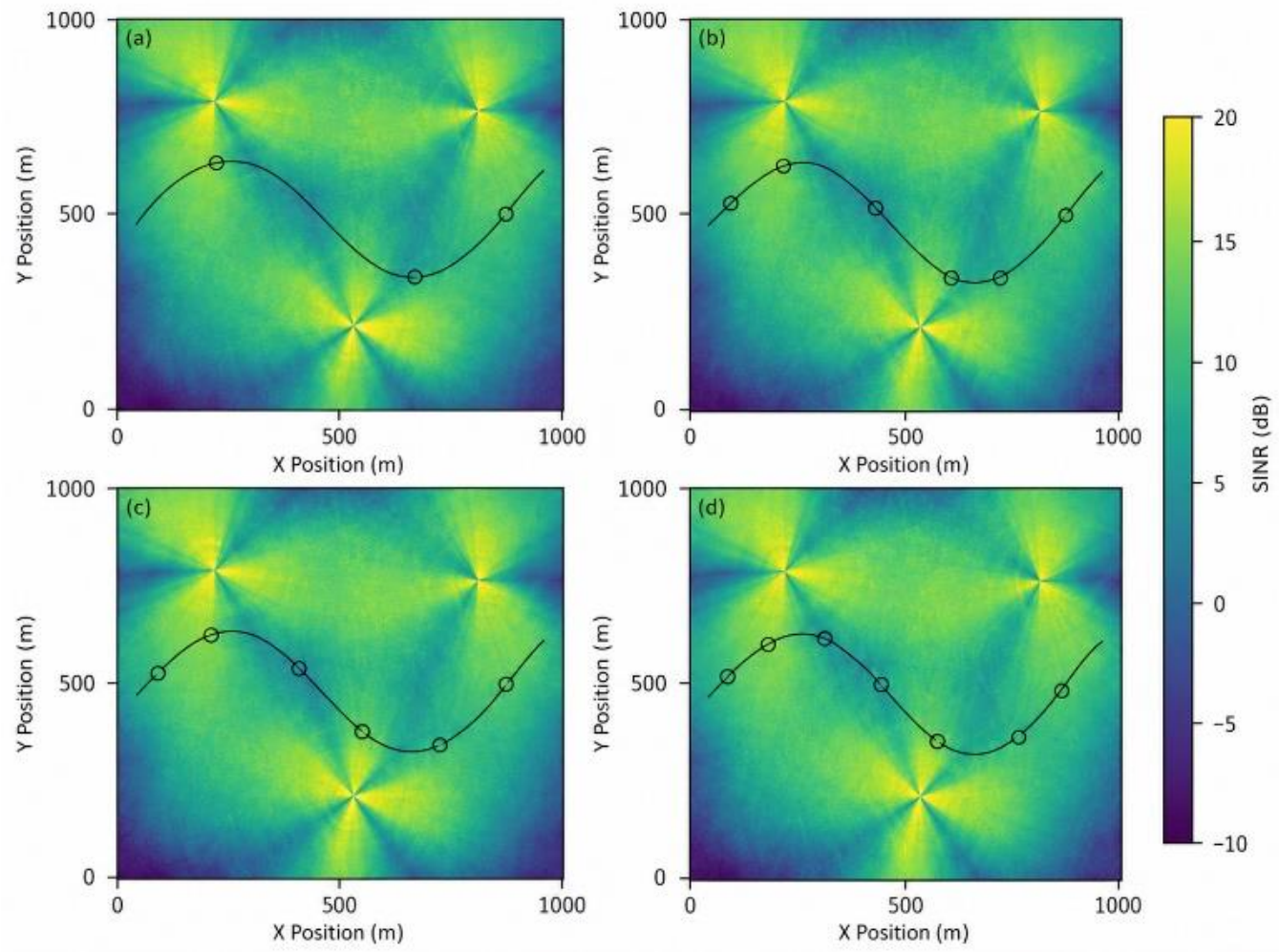

**Figure 27.** SINR heatmap over a 1 km × 1 km macro-cell deployment (tri-sector sites), computed using path loss, log-normal shadowing (σ = 6 dB), and Rayleigh small-scale fading. The same UE trajectory is overlaid in all subplots, while handover locations (×) differ for PPO, CDQL, A3, and ReBuHa, reflecting algorithm-specific association and mobility decisions..

A3 uses a traditional event-based handover algorithm with a certain level of hysteresis and time to trigger, mainly based on the cell that appears to be better in terms of signal quality. ReBuHa also includes a bias based on rules, such as less loaded cells, and hence handovers are shifted earlier or later compared to A3. CDQL uses a learned control behavior that is typically more stepped in nature. Hence, the handovers are shifted accordingly. In PPO, a continuous learned policy is used, and hence it is more likely to be shifted to reduce congestion or improve long-term rewards instead of being based solely on SINR. In summary, it is clear from the figure that although the propagation conditions are the same, there is a significant difference in the points where handovers are made by different handover algorithms.

Figure 28 considers handover behavior from a different perspective: network congestion, as opposed to radio signal quality. The heatmap is a proxy for load/utilization (normalized 0-1), simulating how a real-world macro network would experience uneven traffic and hotspots/congestion issues such as user clustering, traffic bursts, etc. The same user equipment path is included in all subplots to ensure that mobility conditions are identical across all plots. The handover locations are indicative of where each algorithm performed a handover to a new cell, and their size is indicative of the target cell's load at the time of handover. We can use this visualization to determine whether a mobility strategy is attempting to offload a UE to a less loaded cell or a highly loaded cell.

A3 is primarily a signal-driven handover method and does not explicitly account for cell load. As a result, it may trigger handovers toward highly loaded sectors when those sectors offer more favorable radio conditions. In contrast, ReBuHa incorporates load-awareness into its handover decisions and is therefore more responsive to load imbalance. This can shift handovers away from congested regions or increase handover activity around load gradients as the algorithm attempts to relieve overloaded cells. This behavior is reflected by the concentration of handover points near load-transition regions and by the smaller target-cell load markers in those areas.

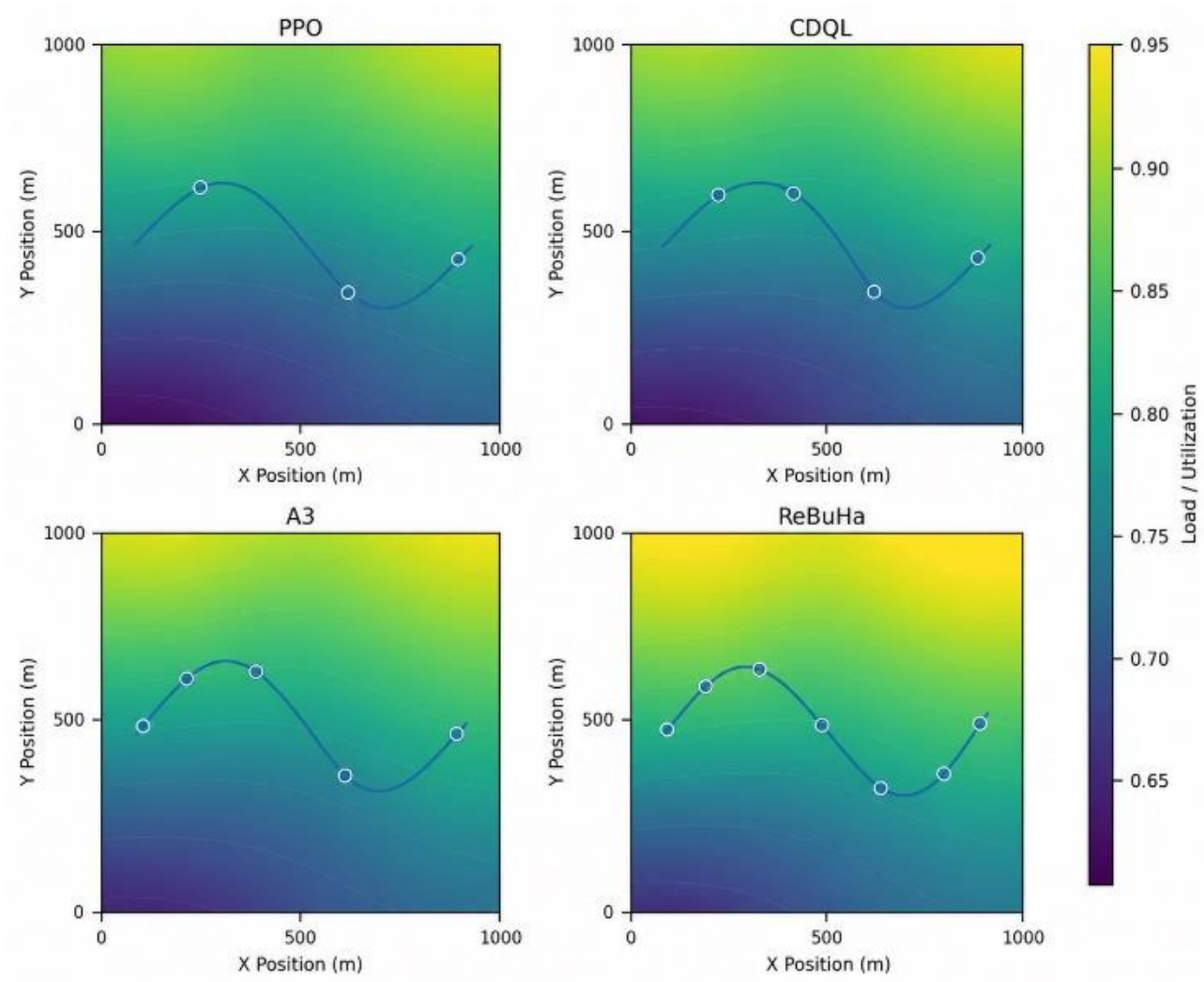

**Figure 28**. Network load/utilization heatmap with the UE trajectory overlaid; algorithm-dependent handover locations are shown, with handover marker size proportional to the target-cell load at the switching instant.

Finally, CDQL is generally more cautious in terms of handovers due to quantization and tuning for stability; it may therefore tolerate some level of congestion in order to avoid unnecessary mobility events and therefore fewer handovers or more delayed offloading. PPO, being a method that optimizes a long-term goal that may include penalties for high loads, always avoids high-load targets but is not reactive. In brief, the figure 28 establishes a direct relationship between mobility events and level of congestion and makes it obvious that the differences in handover locations are not due to chance but rather due to inherent or explicit preferences for balancing radio quality, utilization, and stability.

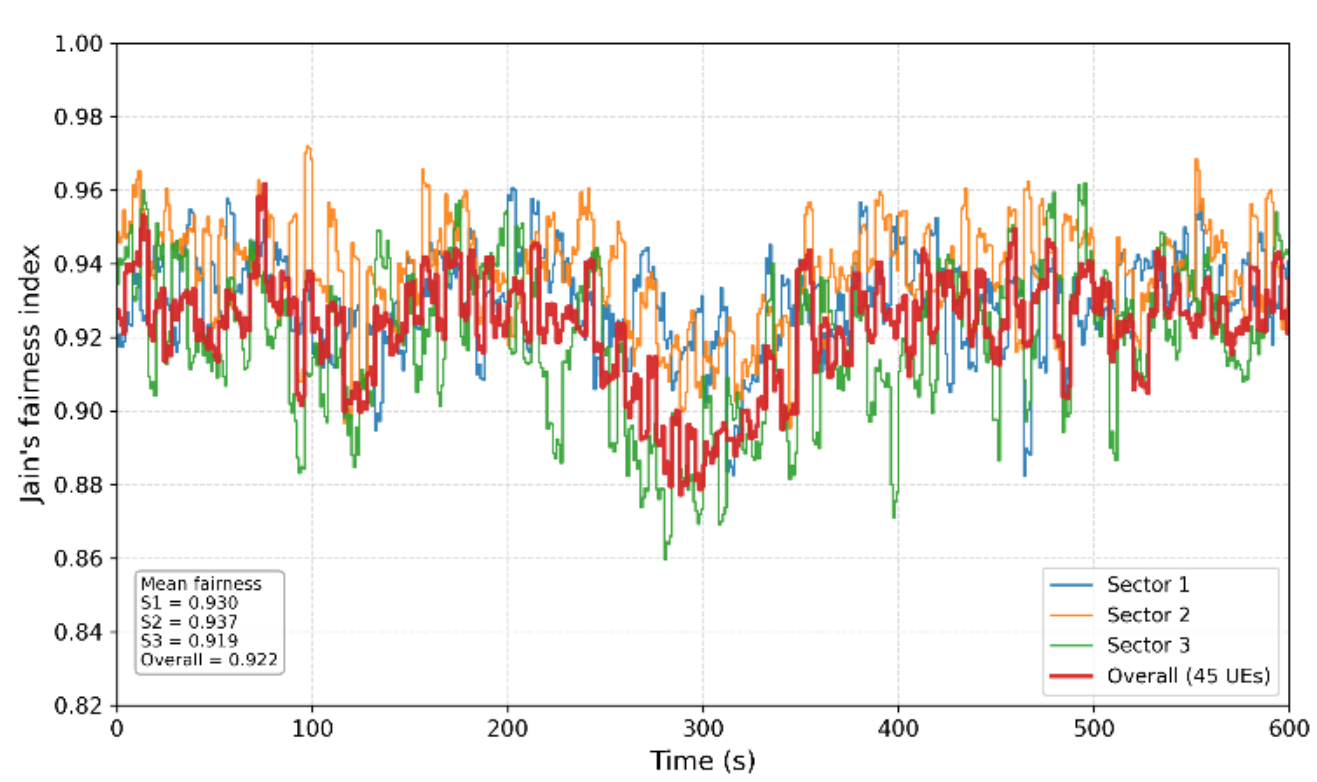

**Figure 29**. Jain's fairness over time for a 45-UE, 3-sector scenario (sector-wise and overall).

Figure 29 tracks how evenly throughput is shared among users during the episode. At each time step, Jain's fairness index is computed from the per-UE downlink throughputs (within each sector for the colored curves, and across all 45 UEs for the black curve). Jain's index ranges from 0 to 1, where 1 means perfectly equal sharing and lower values mean some users are getting much more throughput than others. Here, all three sectors stay mostly in the ~0.90–0.96 range, indicating generally good fairness, while the overall fairness (black) summarizes the network-wide balance and averages around 0.921 (as shown in the annotation). The short-lived dip in the overall curve shows a temporary imbalance—typically caused by a brief period where a subset of users experiences much better radio conditions (or receives more scheduled resources) than others, or when mobility/handovers and traffic bursts create uneven load and scheduling opportunities. The small rapid fluctuations are expected because fairness is sensitive to instant per-UE rate changes (fading, queue dynamics, scheduling decisions). Overall, the plot suggests the system maintains high fairness most of the time, with occasional transient drops when conditions become temporarily uneven.

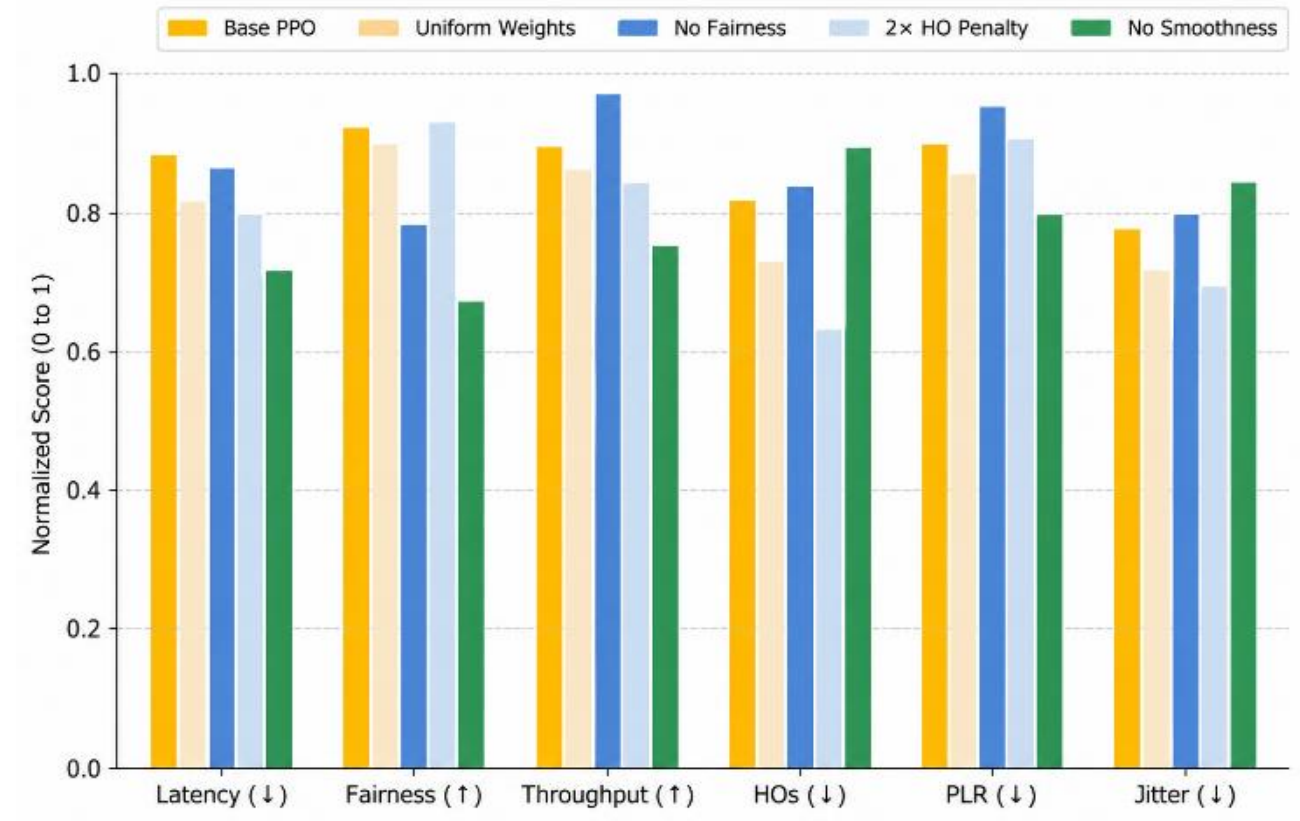

**Figure 30**. PPO reward ablations.

Figure 30 presents the reward-ablation results using normalized KPI scores, where all metrics are mapped to a common [0, 1] scale and higher values indicate better performance. For throughput and fairness, larger values directly represent improvement, while for latency, jitter, PLR, and handovers, the values are inverted utilities; therefore, a higher score means lower delay, lower jitter, lower packet loss, and fewer handovers. This representation makes it easier to compare the effect of each reward component across different QoS objectives.

The Base PPO configuration achieves the most balanced overall performance across the six KPIs. It does not maximize only one objective; instead, it maintains strong throughput and fairness while also keeping latency, jitter, PLR, and handover activity low. This confirms that the selected reward weights provide a reasonable trade-off between network efficiency, user-perceived QoS, and mobility stability.

The Uniform Weights variant performs slightly worse than Base PPO across most metrics. This shows that treating all reward components equally is not optimal for the considered load-balancing problem. Some objectives, such as throughput, fairness, and latency, have a stronger effect on overall QoS and therefore require higher priority in the reward design. Equal weighting can underemphasize these operationally important objectives and lead to a less effective control policy.

The No Fairness variant clearly reduces the fairness score and also affects other QoS metrics. Without the fairness term, the agent has less incentive to distribute load evenly across cells. As a result, it may favor decisions that improve local throughput or short-term performance while allowing imbalance between sectors. This can create overloaded cells, which indirectly increases delay variation, packet loss, and handover instability.

The 2 ×HO Penalty variant improves the handover-related score because the policy becomes more conservative and avoids unnecessary mobility events. However, this comes with a trade-off. When the handover penalty is too strong, the agent may avoid beneficial handovers that could relieve congestion or improve service quality. Therefore, the throughput and some QoS scores can decrease compared with Base PPO. This result shows that handover reduction is useful, but over-penalizing handovers can make the controller too passive.

The No Smoothness variant shows weaker stability-related behavior. Removing the CIO smoothness penalty allows the policy to make more abrupt CIO changes, which can disturb association boundaries and increase oscillatory behavior near cell edges. Although this variant may still perform reasonably in some individual metrics, it provides a less stable and less balanced QoS profile overall. This confirms that smooth CIO updates are important for practical mobility load balancing.

To evaluate the role of each reward and control component, an ablation study was performed by changing one design element at a time while keeping the rest of the training setup unchanged. Table 6 provides a qualitative summary of the ablation trends observed from the normalized KPI scores in Figure 23

Table 6. Qualitative summary of PPO ablation behavior

| **Metric / Observation** | **Base PPO** | **Uniform weights** | **No fairness term** | **2× HO penalty** | **No CIO smoothness** |
|---|---|---|---|---|---|
| **Throughput** | Highest | Lower | Comparable / slightly high | Lower | Unstable |
| **Latency** | Lowest | Higher | Higher | Higher | Higher |
| **PLR** | Lowest | Higher | Higher | Higher | Higher |
| **Jitter** | Lowest | Higher | Higher | Higher | Higher |
| **Fairness** | Highest | Lower | Much lower | Lower | Lower |
| **Handovers** | Low | Similar | Less stable | Lowest | Higher |
| **Key observation** | Best overall QoS balance | Equal weighting weakens QoS prioritization | Load imbalance increases despite acceptable throughput | Over-penalizing handovers limits useful offloading | Abrupt CIO changes cause oscillatory control |

. As summarized in Table 6, the base PPO configuration achieves the best overall QoS trade-off. Uniform weighting reduces performance because it does not reflect the practical priority of key QoS metrics. Removing the fairness term weakens load balancing and increases tail delay and instability. Doubling the handover penalty reduces handovers, but it makes the controller too conservative and prevents useful offloading from congested cells. Removing the CIO smoothness penalty leads to unstable CIO changes, higher jitter, and more handovers. These results confirm that both the multi-objective reward design and the smoothness regularization are necessary for stable QoS-aware load balancing.

## VI. DISCUSSIONS

The results show that the proposed PPO-based load-balancing controller can improve the overall QoS trade-off in the considered 5G cellular network scenario. By applying continuous and bounded CIO adjustments, the agent can influence user association gradually instead of forcing direct handovers. This allows traffic to be shifted from congested cells to neighboring cells when the expected long-term QoS benefit outweighs the cost of changing the serving cell. Compared with A3 and ReBuHa, PPO performs better because it is not limited to fixed thresholds or short-term radio/load indicators. Instead, it learns from multiple QoS metrics, including throughput, latency, jitter, packet-loss ratio, fairness, and handover activity. This helps the controller make more balanced decisions, especially when the strongest signal does not correspond to the best service quality. Compared with CDQL, PPO also benefits from its continuous action space, which is more suitable for smooth CIO control.

Another important result is that PPO reduces unnecessary handovers while maintaining higher throughput and fairness. This suggests that the learned policy does not rely on aggressive user switching. The bounded CIO action and smoothness penalty help prevent abrupt changes in handover bias, leading to more stable mobility behavior near cell boundaries. The scalability results also show that PPO maintains a better QoS trade-off as the number of UEs increases. As expected, heavier load increases latency, jitter, packet loss, and handover activity, while fairness decreases. PPO consistently outperforms CDQL across the evaluated UE-density levels, indicating better congestion-aware load redistribution under dense network conditions. Nevertheless, the results should be interpreted within the limits of the simulation setup. The evaluation is based on a custom Python-based simulator rather than a full 3GPP-compliant system-level simulator or real RAN deployment. Although the simulator includes key elements such as path loss, shadowing, fading, mobility, traffic variation, scheduling, handover logic, and noisy observations, it cannot capture all practical 5G network details. The robustness claims are also limited to the tested uncertainty conditions. More extensive evaluation is needed under severe reporting delay, missing telemetry, trace-driven traffic, larger 7-site and 19-site topologies, and small-cell overlays. In addition, the PPO performance depends on the selected reward weights, which may need to be adapted for different operator priorities, such as low-latency services or throughput-oriented applications. Overall, the results support PPO as a promising approach for QoS-aware CIO control and mobility load balancing. Future work should focus on larger and more realistic network scenarios, reward-weight sensitivity analysis, multi-agent PPO, and O-RAN near-real-time RIC implementation.

## VII. CONCLUSION

This paper presented a PPO-based deep reinforcement learning framework for QoS-aware mobility management and load balancing in 5G cellular networks. The proposed approach formulates CIO-based user association control as a Markov Decision Process, where the agent observes aggregated radio, load, and QoS indicators and learns to generate bounded CIO adjustment actions for each cell. By using CIO as an indirect control variable, the proposed method influences handover decisions and redistributes traffic across neighboring cells while remaining compatible with conventional mobility-management procedures.

The reward function was designed to capture the multi-objective nature of 5G load balancing. It jointly considers aggregate throughput, latency, jitter, packet-loss ratio, Jain's fairness index, and handover activity, together with a smoothness penalty that discourages abrupt CIO variations. This design allows the agent to improve network efficiency while maintaining user-perceived QoS and mobility stability. The PPO agent was implemented with an actor–critic architecture and trained in a Python-based simulation environment that includes Gauss–Markov user mobility, stochastic traffic dynamics, channel variation, QoS-aware scheduling, handover logic, and observation noise.

Simulation results demonstrated that the proposed PPO controller consistently improves the overall QoS trade-off compared with classical A3 and ReBuHa baselines, as well as the learning-based CDQL method. The learned policy increased throughput and fairness while reducing latency, jitter, packet loss, and unnecessary handovers. Under increasing user density, PPO maintained more stable performance and showed better scalability than the baseline approaches. These results show that clipped policy updates and advantage-based learning help produce smoother convergence and more stable CIO-control behavior under mobility and noisy observations.

The proposed framework shows that policy-gradient reinforcement learning can be an effective tool for self-optimizing RAN control, especially when the control objective involves multiple competing KPIs. Rather than relying on fixed handover thresholds or manually tuned load-balancing rules, the PPO-based controller learns adaptive CIO adjustments from network feedback and responds to time-varying traffic, mobility, and congestion conditions.

Despite these promising results, the present study has several limitations. The evaluation is based on a custom Python simulation environment rather than a full 3GPP-compliant system-level simulator or real RAN deployment. In addition, the current formulation considers a single centralized PPO controller for macro-cell load balancing. Future work will focus on validating the method using trace-driven RSRP/CQI and traffic data, extending the framework to multi-agent PPO for distributed inter-cell coordination, and integrating the controller into an O-RAN near-real-time RIC architecture as an xApp. Further extensions may include energy-aware cell activation, backhaul-aware load balancing, safe reinforcement learning constraints, and domain-randomization techniques to improve sim-to-real transfer.